\documentclass[journal]{IEEEtran}
\def\doublecolumn{1}

\usepackage[cmex10]{amsmath}  
\usepackage{amssymb}
\usepackage{graphicx,color}   
\usepackage{mathtools}
\usepackage{cases}    

\usepackage{tikz}
\usetikzlibrary{fit}
\usetikzlibrary{graphs}
\usetikzlibrary{positioning}

\usepackage[T1]{fontenc}
\usepackage{microtype}
\graphicspath{{../pictures/}}

\usepackage{cite}  
\usepackage[caption=false,font=footnotesize]{subfig} 
\usepackage[ruled,vlined]{algorithm2e}

\newtheorem{theorem}{Theorem}
\newtheorem{remark}{Remark}
\newtheorem{lemma}{Lemma}
\newtheorem{definition}{Definition}
\newtheorem{example}{Example}
\newtheorem{proposition}{Proposition}
\newtheorem{corollary}{Corollary}

\newcommand{\total}{{N_\text{leaf}}}
\newcommand{\predecessor}{{V_\text{predec}}}
\newcommand{\connected}{{N_\text{msg-conn}}}

\newcommand{\tree}{{N_\text{tree}}}
\newcommand{\iteration}{{N_\text{iteration}}}
\newcommand{\Soutside}{\mathcal{V}_\text{S}^\text{outside}}
\newcommand{\Sinside}{\mathcal{V}_\text{S}^\text{inside}}

\renewcommand{\S}{\mathcal{V}_\text{S}} 

\newcommand{\remain}{\mathcal{V}^\text{remaining}}

\DeclareMathOperator*{\argmin}{\arg\!\min}

\IEEEoverridecommandlockouts
\begin{document}

\title{\huge The Single-Uniprior Index-Coding Problem:\\ The Single-Sender Case and The Multi-Sender Extension}
\author{Lawrence Ong, Chin Keong Ho, Fabian Lim
\thanks{Part of the material in this paper was presented at
the IEEE International Symposium on Information Theory, Istanbul, Turkey, July 7--12, 2013, and at the IEEE International Conference on Communications, Ottawa, Canada, June 10--15, 2012.

This work was done when Fabian Lim was with Massachusetts Institute of Technology under the support of NSF grant ECCS-1128226.

Lawrence Ong is the recipient of an Australian Research Council Future Fellowship (FT140100219).}}
\maketitle

\begin{abstract}
Index coding studies multiterminal source-coding problems where a set of receivers are required to decode multiple (possibly different) messages from a common broadcast, and they each know some messages a priori. 
In this paper, at the receiver end, we consider a special setting where 
each receiver knows only one message a priori, and each message is known to only one receiver.
At the broadcasting end, we consider a generalized setting where there could be multiple senders, and each sender knows a subset of the messages. The senders collaborate to transmit an index code.
This work looks at minimizing the number of total coded bits the senders are required to transmit. 
When there is only one sender, we propose a pruning algorithm to find a lower bound on the optimal (i.e., the shortest) index codelength, and show that it is achievable by linear index codes. 
When there are two or more senders,  we propose an appending technique to be used in conjunction with the pruning technique to give a lower bound on the optimal index codelength; we also derive an upper bound based on cyclic codes. While the two bounds do not match in general, for the special case where no two distinct senders know any message in common, the bounds match, giving the optimal index codelength. 
The results are expressed in terms of strongly connected components in directed graphs that represent the index-coding problems.
\end{abstract}

\section{Introduction}

We investigate a broadcast problem over noiseless channels with receiver side information, also known as index coding~\cite{baryossefbirk11}. In the classical setup, \textit{one} sender encodes a collection of messages, and broadcasts the codeword to multiple receivers. Each receiver knows some messages a priori, and is to decode a set of messages it wants from the single codeword broadcast by the sender. The aim is to find the shortest codeword that the sender needs to broadcast to ensure that all receivers can decode the messages they want. The index-coding problem remains open to date; only a small number of special cases have been solved~\cite{baryossefbirk11,neelytehranizhang12,arbabjolfaei13,blasiakkleinberglubertzky13,ongisit14,yuneely14}.

In this paper, we consider a class of the index-coding problem, which we refer to as \textit{single-uniprior}, where each receiver knows only one message a priori, but may request multiple messages, and each message is known to only one receiver. 
When there is only \textit{one} sender, we completely solve the single-uniprior index-coding problem. We show that linear index codes are optimal for this class---although linear index codes are suboptimal in general~\cite{lubertzkystav09}, e.g., when two receivers know the same message.

We then extend the single-uniprior problem with a single sender to multiple senders, where each sender knows a subset of the messages. We derive lower and upper bounds on the shortest index codelength, and identify cases where the bounds match. In particular, the bounds match when no two senders know any message in common.

\subsection{Motivation of the multi-sender single-uniprior index-coding problem}

The single-uniprior index-coding problem formulation is motivated by satellite communications~\cite{wynerwolf02,gunduzyener09,ongmjohnsonit11}, where multiple clients exchange messages through a satellite, which acts as a relay. 
Having no direct communication links, the clients first send their messages to the satellite via an \textit{uplink} channel. The satellite decodes the messages, re-encodes, and broadcasts to the clients via a \textit{downlink} channel. 
Here, the downlink corresponds to a single-uniprior index-coding problem, where each client is to obtain messages of other clients, and it knows only its own message a priori. 

When multiple satellites are used, the downlink corresponds to a multi-sender index-coding problem. In this case, different senders know different messages a priori, due to, for instance, decoding errors when getting the messages from the clients on the uplink, line-of-sight only to some clients, or limited storage of the senders.
To the best of our knowledge, this is the first paper that investigates the multi-sender index-coding problem. 

\section{Channel Model and Notation} \label{section:model}

An \textit{instance} of the multi-sender index-coding problem (including the single-sender as a special case) consists of the following:
\begin{itemize}
\item $m$ independent messages, denoted by an ordered\footnote{\label{note1}The elements are ordered in increasing indices.} set $\mathcal{M} = \{x_1, x_2, \dotsc, x_m\}$.
\begin{itemize}
\item Each message $x_i$ has $q_i$ binary bits, denoted by $x_i = [x_i[1], x_i[2], \dotsc, x_i[q_i]]$, where each bit $x_i[\cdot]$ is drawn from an independent and uniform distribution over $\{0,1\}$.
\end{itemize}
\item $S$ senders: each sender $s \in \{1,2,\dotsc, S\}$ knows a subset of the messages, denoted by an ordered\textsuperscript{\ref{note1}} subset $\mathcal{M}_s \subseteq \mathcal{M}$.
\item $n$ receivers: each receiver~$r \in \{1,2,\dotsc, n\}$ 
\begin{itemize}
\item knows an ordered\textsuperscript{\ref{note1}} subset of the messages a priori, $\mathcal{K}_r \subset \mathcal{M}$, and
\item wants/requests an ordered\textsuperscript{\ref{note1}} subset of the messages, $\mathcal{W}_r \subseteq \mathcal{M}$.
\end{itemize}
\end{itemize}
Figure~\ref{fig-0} depicts an example of a multi-sender \textit{single-uniprior} index-coding instance, where $\mathcal{K}_i = \{x_i\}$.

Without loss of generality, we assume that $\mathcal{W}_r \cap \mathcal{K}_r = \emptyset$ for any $r$, and that $\bigcup_{s=1}^S \mathcal{M}_s = \mathcal{M}$, i.e., each message is available at some sender(s). 

\ifx\doublecolumn\undefined
\begin{figure}[t]
\centering
\includegraphics[width=10cm]{mwrc-it-09-a}
\caption{The multi-sender single-uniprior index-coding problem: We consider distributed-like settings where each sender $s$ is limited to know a subset $\mathcal{M}_s$ of the message set $\mathcal{M}$. Also, we consider a multicast setup where each receiver~$r$ knows a single unique message $x_r$ a priori, and wants a subset of messages $\mathcal{W}_r \subset \mathcal{M}$. Each sender $s$ broadcasts $\boldsymbol{c}_s$ to all receivers noiselessly. The aim is to find the minimum aggregated codelength $|\boldsymbol{c}_1|+|\boldsymbol{c}_2|+|\boldsymbol{c}_3|$.}
\label{fig-0}
\end{figure}
\else
\begin{figure}[t]
\centering
\includegraphics[width=8.5cm]{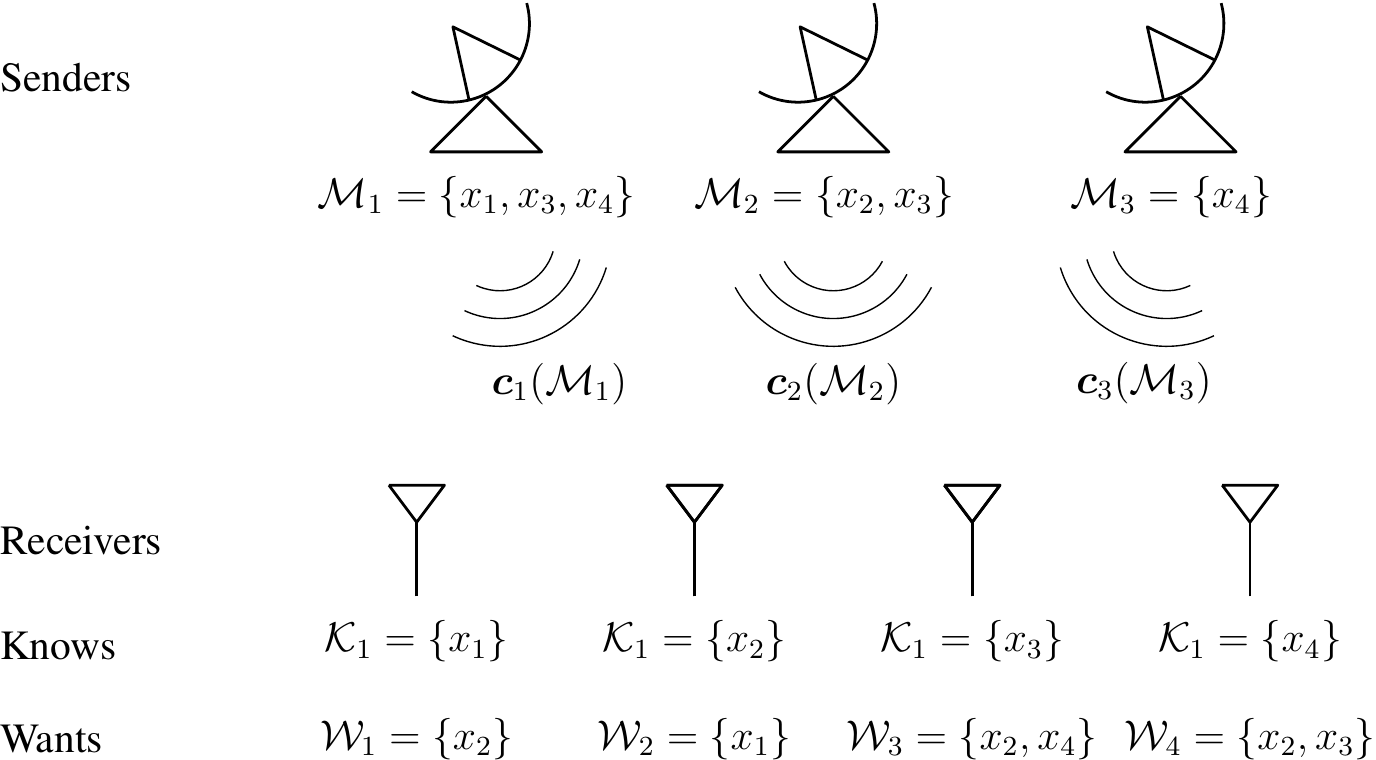}
\caption{The multi-sender multicast single-uniprior index coding: We consider distributed-like settings where each sender $s$ is limited to know a subset $\mathcal{M}_s$ of the message set $\mathcal{M}$. Also, we consider a multicast setup where each receiver~$r$ knows a single unique message $x_r$ a priori, and wants a subset of messages $\mathcal{W}_r \subset \mathcal{M}$. Each sender $s$ broadcasts $\boldsymbol{c}_s$ to all receivers noiselessly. The aim is to find the minimum aggregated codelength $|\boldsymbol{c}_1|+|\boldsymbol{c}_2|+|\boldsymbol{c}_3|$.}
\label{fig-0}
\end{figure}
\fi

We define a multi-sender index code for the above setup:
\begin{definition}[Multi-Sender Index Code]
An index code for an index-coding instance consists of
\begin{enumerate}
\item an encoding function for each sender $s \in \{1,2,\dotsc, S\}$,
\begin{equation*}
\mathtt{E}_s: \{0,1\}^{\sum_{i:x_i \in \mathcal{M}_s} q_i} \mapsto \{0,1\}^{\ell_s},
\end{equation*}
\item a decoding function for each receiver~$r \in \{1,2,\dotsc, n\}$,
\begin{equation*}
\mathtt{D}_r: \{0,1\}^{\sum_{s=1}^S \ell_s +\sum_{i:x_i \in \mathcal{K}_r} q_i} \mapsto \{0,1\}^{\sum_{i:x_i \in \mathcal{W}_r}q_i},
\end{equation*}
\end{enumerate}
such that $\mathcal{W}_r = \mathtt{D}_r(\mathtt{E}_1 (\mathcal{M}_1), \mathtt{E}_2 (\mathcal{M}_2), \dotsc, \mathtt{E}_S (\mathcal{M}_S), \mathcal{K}_r)$ for each receiver~$r$.
\end{definition}

This means each sender~$s$ encodes its known messages into an $\ell_s$-bit sub-codeword. The sub-codewords of all senders are given to all receivers.
The total number of transmitted bits is $\tilde{\ell} \triangleq \sum_{s=1}^S \ell_s$, which is the index codelength. 
We seek the optimal (i.e., the minimum) index codelength, denoted by $\tilde{\ell}^*$, as well as  index codes with the optimal length.  
If there is only one sender, $S=1$, we simplify the notation by denoting the optimal index codelength as $\ell^*$.

The case $S=1$ reduces to the single-sender index-coding problem studied in many works~\cite{
birkkol2006,
alonhassidim08, 
lubertzkystav09,
rouayhebsprintsongeorghiades10,
baryossefbirk11,
chaudhry11,
ongho12,
tehranidimakisneely12,
dauskachekchee12,
neelytehranizhang12,
arbabjolfaei13,
blasiakkleinberglubertzky13,
shanmugamdimakislangberg13,
shanmugamdimakis14,
arbabjolfaeikim14,
yuneely14,
ebrahimisiavoshani14, 
ongnetcod14,
ongisit14,
arbabjolfaeikim15,
thapaongjohnson15}.
Clearly, for the same receiver setting $\{\mathcal{W}_r, \mathcal{K}_r\}_{r=1}^n$, a multi-sender index code for any $S > 1$ case will also be a code for the single-sender $S=1$ case.
This is because, by definition, the only sender in the single-user case, who knows all the messages, can transmit any multi-sender index code.
But the converse is not true. 
So, techniques for the single-sender case do not straightforwardly apply to the multi-sender case.

For both the single-sender and the multi-sender index-coding problems, we assume that each sender knows which messages (not the message content) each receiver knows, $\mathcal{K}_r$, and requests, $\mathcal{W}_r$; and which  messages each sender has, $\mathcal{M}_s$. This can be achieved by having the nodes to communicate (the message-availability and message-request information) prior to index coding. Under such scenarios, index coding is useful if the messages are sufficiently long such that the overhead incurred during the prior communication is negligible.\footnote{For simplicity, in the multi-sender section of this paper, we assume that each message consists of a single bit. However, all proofs are also valid even when each message consist of $q$ bits, for any positive integer $q$.} In some communication scenarios, side information is strategically placed at the receivers (for example, data pre-fetching or caching~\cite{maddahalineisen14}), and the messages in the senders' storage are populated according to their geographical locations. In such cases, the senders can be informed of all message indices in $\mathcal{K}_r$, $\mathcal{W}_r$, and $\mathcal{M}_s$.

If we set the number of senders to one, $S=1$, and the messages to be of equal length, $q_i = q$ for all $i$, the optimal \textit{normalized} index codelength---over all index codes and all message lengths---is commonly known as the \textit{optimal broadcast rate}~\cite{blasiakkleinberglubertzky13}:
\begin{equation}
\beta \triangleq \inf_{q} \frac{\ell^*}{q} = \lim_{q \rightarrow \infty} \frac{\ell^*}{q} \text{ codeword length per message length}.
\end{equation}
The second equality is due to the sub-additivity property of $\ell^*$~\cite{alonhassidim08}.

\subsection{Terminology}

We will represent index-coding instances by graphs, and derive bounds on $\ell^*$ by executing graph operations.
We will use common graph terminology:
A \textit{directed path} of length $L$ from vertex $v_0$ to vertex $v_L$ consists of the following $(L+1)$ distinct vertices (except possibly the first
and last) and $L$ arcs: $\{ v_0, (v_0 \rightarrow v_1), v_1, (v_1 \rightarrow v_2), \dotsc, (v_{L-1} \rightarrow v_L), v_L\}$. A \textit{cycle} is a path with the same first and last vertices.
A \textit{strongly connected component} (SCC) of a graph is a \textit{maximal} subgraph such that for any vertex pair $(i,j)$ in the subgraph, there is a directed path in the subgraph from $i$ to $j$, and another from $j$ to $i$. A \textit{trivial SCC} is an SCC with only one vertex.
A \textit{leaf} vertex in a graph has no outgoing arcs. Similarly, a \textit{leaf non-trivial SCC} is an SCC with two or more vertices, in which there is no arc from any vertex in the SCC to any vertex outside the SCC. In other words, for each vertex in the SCC, its out-neighbors are all vertices in the SCC.
A vertex $j$ is a \textit{predecessor} of vertex $i$ if and only if there is a directed path from $j$ to $i$.
A \textit{path} on an undirected graph is similarly defined with arcs replaced by edges. A subgraph is \textit{connected} if there is a path within the subgraph from any vertex to any other vertex. A \textit{tree} is a connected undirected subgraph with no cycle.


\section{Index-Coding Classification and Graphical Representation}

A single-sender instance is fully described by the message lengths $\{q_i\}_{i=1}^m$ and the receiver setting $\{\mathcal{W}_r, \mathcal{K}_r\}_{r=1}^n$. For the multi-sender extension, we further require the sender setting $\{\mathcal{M}_s\}_{s=1}^S$.

\subsection{Classification of index-coding instances based on the receiver setting}

We first consider the single-sender setup, and categorize the instances as follows:

\subsubsection{Information flow}
We classify different types of information flow from the sender to the receivers. We say that an index-coding instance is \textit{unicast} if

\begin{equation}
\mathcal{W}_i \cap \mathcal{W}_j = \emptyset, \quad \forall i \neq j, \label{eq:unicast}
\end{equation}
meaning that each message can be requested by at most one receiver.
 In addition, we say that the instance is \textit{single-unicast} if, in addition to \eqref{eq:unicast}, we also have that $|\mathcal{W}_i| = 1$ for all $i$, meaning each receiver requests exactly one message---but the message may consist of multiple bits.

Note that any unicast instance can be recast as a single-unicast instance~\cite{rouayhebsprintsongeorghiades10}. Suppose that a receiver~$i$ wants two messages $\mathcal{W}_i = \{x_1, x_2\}$, and knows $\mathcal{K}_i$. 
As far as index codes are concerned, we can replace receiver~$i$ by two receivers $i'$ and $j'$, both knowing the same set $\mathcal{K}_i$, and each wanting a single message, i.e., $x_1$ and $x_2$ respectively.  
We can further split the receivers in a similar manner so that each message contains only a single bit.

Unicast instances were investigated by Bar-Yossef et al.~\cite{baryossefbirk11}, Lubetzky and Stav~\cite{lubertzkystav09}, Shanmugam et al.~\cite{shanmugamdimakislangberg13,shanmugamdimakis14}, Arbabjolfaei et al.~\cite{arbabjolfaei13,arbabjolfaeikim14,arbabjolfaeikim15}, Yu and Neely~\cite{yuneely14}, Wong et al.~\cite{wonglangberg13}, Ebrahimi and Siavoshani~\cite{ebrahimisiavoshani14}, and Ong et al.~\cite{ongnetcod14,ongisit14,thapaongjohnson15}.

\subsubsection{Side information}
We next classify different types of side information at the receivers. We say that an index coding instance is \textit{uniprior} if

\begin{equation}
\mathcal{K}_i \cap \mathcal{K}_j = \emptyset, \quad \forall i \neq j, \label{eq:uniprior}
\end{equation}
meaning that each message is known a priori to at most one receiver. In addition, we say that the instance is \textit{single-uniprior} if, in addition to \eqref{eq:uniprior}, we also have that $|\mathcal{K}_i| = 1$ for all $i$, meaning that each receiver knows exactly one unique message a priori. 

Unlike unicast instances, single-uniprior instances do not subsume all uniprior instances. A receiver who knows $\{x_1, x_2\}$ is not equivalent to two receivers each knowing only one of them.

Unicast uniprior instances were investigated by Neely et al.~\cite{neelytehranizhang12}.

With the above terminology, we call index-coding instances with no restrictions on $\mathcal{W}_i$ and $\mathcal{K}_i$  \textit{multicast multiprior} instances. 
Multicast multiprior instances were investigated by Blasiak et al.~\cite{blasiakkleinberglubertzky13}, Alon et al.~\cite{alonhassidim08}, Shanmugam et al.~\cite{arbabjolfaeikim14}, Tehrani et al.~\cite{tehranidimakisneely12}, and Neely et al.~\cite{neelytehranizhang12}.



%

\subsection{Graphical representation of the receiver setting}

Existing works on index coding focused on the single-sender case, and many used the following graphical representations to capture the receivers' side information and requests.

\subsubsection{Unicast multiprior (or simply unicast)}
As mentioned above, any unicast instance can be converted into an equivalent single-unicast instance. 
Without loss of generality, we can assume exactly $n$ receivers and $n$ messages, as message not requested by any receivers can be removed.
Single-unicast index-coding instances are commonly represented by \textit{side-information graphs}~\cite{baryossefbirk11} with $n$ vertices $\{1,2,\dotsc,n\}$, where an arc exists from vertex $i$ to vertex $j$ if and only if receiver~$i$ knows the message requested by receiver~$j$. 


\begin{figure}[t]
\centering
\subfloat[Side-information graph $\mathcal{G}_1$]{
\qquad\quad
\resizebox{9ex}{!}{%
\begin{tikzpicture}
\graph {[grow right=1.5cm, branch down=1.5cm, nodes={draw,circle},edge={>=latex}]
1 -!- 4, 
3 -!- 2,
1 ->[bend right] 4,
1 -> 3,
2 -> 4, 
2 ->[bend right] 3,
4 ->[bend right] 1,
3 ->[bend right] 2,
};
\end{tikzpicture}
}
\qquad\quad
}
\qquad
\subfloat[Side-information graph $\mathcal{G}_2$]{
\resizebox{23ex}{!}{%
\begin{tikzpicture}
\graph {[nodes={draw,circle},edge={>=latex} ,clockwise = 4, chain polar shift=(0:1.5cm), radius=1cm]
4, 2 ->[bend right] 3, 5, 1 ->[bend right] 6,
6 ->[bend right] 1,
3 ->[bend right] 2,
1 -> 5 -> 2 -> 4 -> 1
};
\end{tikzpicture}
}%
}
\caption{Incorrect side-information graphs for the uniprior instance in Example~\ref{ex:unicast}}
\label{fig:unicast-example}
\end{figure}

\subsubsection{Unicast uniprior}

The class of index-coding instances with unicast and uniprior was investigated by Neely et al.~\cite{neelytehranizhang12}, where each message bit is (i) known to only one receiver, and (ii) requested by only one receiver. 
Neely et al.\ represented this class by a weighted compressed graph, where each vertex represents a receiver, and an arc of weight $q$ exists from vertex $i$ to $j$ if and only if receiver~$j$ wants $q$ messages known to receiver~$i$.

Although a unicast instance can be recast into a single-unicast instance, not all unicast uniprior instances can be recast into single-unicast uniprior instances. This is because if we split a receiver, there will be more than one receivers knowing the same message (i.e., the instance is no longer uniprior).

\subsubsection{Multicast multiprior}

This is the most general class of the index-coding instances, where there is no restriction on $\mathcal{W}_i$ and $\mathcal{K}_i$.
To completely capture the information in both $\mathcal{W}_i$ and $\mathcal{K}_i$, Neely et al.~\cite{neelytehranizhang12} proposed a bipartite-graph representation, where there are $n$ receiver vertices, $\{r_1, r_2, \dotsc, r_n\}$, and $m$ messages vertices, $\{x_1, x_2, \dotsc, x_m\}$. An arc from $r_i$ to $x_j$ exists if and only if receiver~$i$ wants $x_j$, and arc from $x_i$ to $r_j$ exists if and only if receiver~$j$ knows $x_i$. Blasiak et al.~\cite{blasiakkleinberglubertzky13} and Alon et al.~\cite{alonhassidim08} used hypergraphs to represent multicast multiprior instances.

\subsubsection{Multicast single-uniprior (or simply single-uniprior)}

This is the class of index-coding instances considered in this paper.
Without loss of generality, we consider $n$ receivers and $n$ messages. This subsumes the case where a message $x_j$ is not known to any receiver. To see this, add a dummy receiver~$n+1$ who knows $x_j$ but does not want any message. Clearly,  adding this receiver does not change the problem.

Side-information graphs used for single-unicast instances cannot capture all single-uniprior instances. We illustrate this using the example below:
\begin{example} \label{ex:unicast}
Consider the following uniprior instance:
$ 
\mathcal{K}_1 = \{x_2\}, \mathcal{K}_2 = \{x_1\}, \mathcal{K}_3 = \{x_4\}, \mathcal{K}_4=\{x_3\}, 
\mathcal{W}_1 = \{x_3\}, \mathcal{W}_2 = \{x_4\}, \mathcal{W}_3= \{x_1,x_2\}, \mathcal{W}_4 = \{x_1, x_2\}, 
$
where each message $x_i$ contains one bit. We will see later (Corollary~\ref{corollary:single-sender-binary}) that the optimal index codelength for this instance is three bits.
If we apply the rule for constructing side-information graphs that says ``$i \rightarrow j$ exists if receiver~$i$ knows the message requested by receiver~$j$'', we get graph $\mathcal{G}_1$ shown in Figure~\ref{fig:unicast-example}. For a unicast instance represented by $\mathcal{G}_1$, the optimal codelength is two bits (using clique cover~\cite{birkkol2006}). Thus, $\mathcal{G}_1$ incorrectly represents the uniprior instance described here. However, since receivers~3 and 4 request multiple messages, we can obtain an equivalent instance by splitting receiver~3 to two receivers (say, 3 and 5), each knowing the same message but requesting a different message: $\mathcal{K}_3 = \{x_4\}, \mathcal{K}_5 = \{x_4\}, \mathcal{W}_3 = \{x_1\}, \mathcal{W}_5 = \{x_2\}$. We similarly split receiver~4: $\mathcal{K}_4=\{x_3\}, \mathcal{K}_6 = \{x_3\}, \mathcal{W}_4 = \{x_1\}, \mathcal{W}_6 = \{x_2\}$. Constructing a side-information graph for this equivalent instance,   we get $\mathcal{G}_2$ in Figure~\ref{fig:unicast-example}. But for a unicast instance represented by $\mathcal{G}_2$, the optimal codelength is four bits (again, using clique cover). Thus $\mathcal{G}_2$ again incorrectly represents the uniprior instance in this example. The reason is that side-information graphs are not able to capture multiple receivers requesting a same message.

\end{example}

Instead, we propose \textit{information-flow graphs} with $n$ \textit{weighted} vertices.
Without loss of generality, let $\mathcal{K}_i = \{x_i\}$ for all receiver~$i \in \{1,2,\dotsc,n\}$.
An arc exists from vertex $i$ to vertex $j$ if and only if node $j$ wants the message known to receiver~$i$, i.e., $x_i \in \mathcal{W}_j$.
The weight of vertex $i$ is $q_i$, which is the number of bits in message $x_i$. 
Although a uniprior instance can be represented by either a hypergraph or a bipartite graph, using the information-flow graph results in a smaller (compare to the bipartite graph) and simpler (a directed graph instead of a hyper graph) graph.

\subsection{Graphical representation of the sender setting} \label{section:multi-sender-ext}

When there are multiple senders, we need to represent $\{\mathcal{M}_s\}$ in addition to $\{q_i, \mathcal{K}_r, \mathcal{M}_r\}$. We propose to represent $\{\mathcal{M}_s\}$ using an $n$-vertex undirected \textit{message} graph, denoted by $\mathcal{U}$. 
An edge between vertices $i$ and $j$ on $\mathcal{U}$ (denoted by $(i,j)$) exists if and only if messages $x_i$ and $x_j$ are known to the same sender, i.e., $i,j \in \mathcal{M}_s$ for some $s$. 

In this paper, we will consider multi-sender single-uniprior instances each described by an information-flow graph $\mathcal{G}$ and a message graph $\mathcal{U}$, both defined on the same set of vertices $\mathcal{V} = \{1,2,\dotsc, n\}$.
We denote by $\tilde{\ell}^*(\mathcal{G},\mathcal{U})$ the optimal index codelength for the instance represented by $(\mathcal{G},\mathcal{U})$.

Note that an edge $(i,j)$ in $\mathcal{U}$ does not indicate which sender(s) owns both the messages (i.e. the set of $s$ such that $i,j \in\mathcal{M}_s$).
This ambiguity will not affect the techniques developed in this paper. In fact, for some cases, this is sufficient to derive the optimal index codelength. However, we will point out in the conclusion the existence of an instance where
resolving this ambiguity can further improve our results.

\section{Existing Results and Our Contributions}

In this section, we will survey existing results on $\ell^*$ for the single-sender problem. We will then review related works on multiterminal networks with side information that have settings similar to that in the multi-sender index-coding problem. Lastly, we present the main results of this paper.

\subsection{Existing results for single-sender index coding}


Recall that any unicast instance can be first converted to an equivalent single-unicast instance with binary messages and then represented by a side-information graph. For binary messages, Bar-Yossef et al.~\cite{baryossefbirk11} found the optimal index codelength $\ell^*$ for all acyclic side-information graphs. Suppose that a side-information graph has cycle(s) and is \textit{symmetrical} (meaning that for every arc from vertex $i$ to vertex $j$, there is another arc from vertex $j$ to vertex $i$). For this special case, the directed side-information graph can be converted to an undirected graph where an edge between vertices $i$ and $j$ represent the arcs in both directions. Bar-Yossef et al.\ solved all unicast index-coding instances with the following undirected side-information graphs: (i) perfect graphs, (ii) odd holes with five or more vertices, and (iii) odd anti-holes with five or more vertices.

Arbabjolfaei et al.~\cite{arbabjolfaei13} found $\ell^*$ for all unicast instances up to five receivers. They used random-coding arguments to prove achievability. Later, Ong~\cite{ongisit14} showed that binary linear codes are sufficient to achieve $\ell^*$ for all unicast instances up to five receivers. Blasiak et al.~\cite{blasiakkleinberglubertzky13} and Thapa et al.~\cite{thapaongjohnson15} found $\ell^*$ for certain special classes of graphs.

Also for unicast instances, using the bipartite-graph representation, Yu and Neely~\cite{yuneely14} showed that if the graph is planar,\footnote{A graph is planar if it can be drawn on a two-dimensional plane in such a way that its edges intersect only at their endpoints.} then $\ell^*$ can be found using linear programming (solutions not in closed form). This class of unicast instances subsume all unicast uniprior instances up to four receivers as special cases.
Neely et al.~\cite{neelytehranizhang12} found $\ell^*$ for unicast uniprior instances if the weighted compressed graph contains only arc-disjoint cycles. 

For the most general multicast multiprior case, Neely et al.~\cite{neelytehranizhang12} found $\ell^*$ for all acyclic bipartite graphs.


It has been shown~\cite{chaudhry11} that (i) the general multicast multiprior index-coding problem is NP-hard, and (ii) the multicast (non-unicast) index-coding problem is even NP-hard to approximate.

For all the above-mentioned classes of index-coding instances where $\ell^*$ has been found,
\textit{linear} index codes are optimal. However, for some single-unicast instances, Lubetzky and Stav~\cite{lubertzkystav09} showed that non-linear index codes can outperform linear codes.




\begin{table*}
\centering
\begin{tabular}{c c c c l c} 
Sender(s) & Bound & Section & Theorem & Techniques & Bound tight?\\
\hline
single & lower & \ref{sec:single-upper} & Theorem~\ref{theorem:single-sender-lower} & pruning & yes\\
single & upper &  \ref{sec:single-lower} & Theorem~\ref{theorem:single-sender-upper} & cyclic codes on leaf SCCs & yes\\
multiple & lower &  \ref{sec:multiple-lower} & Theorem~\ref{theorem:multi-sender-lower-bound} & pruning and appending & for some cases\\
multiple & upper & \ref{section:achievable} & Theorem~\ref{theorem:multi-sender-achievable} & cyclic codes on leaf SCCs and trees & for some cases
\end{tabular}
\vspace*{1ex}
\caption{A summary of our techniques used to obtain different bounds in this paper}
\label{table:result}
\end{table*}

\begin{figure*}[t]
\centering
\includegraphics[width=14cm]{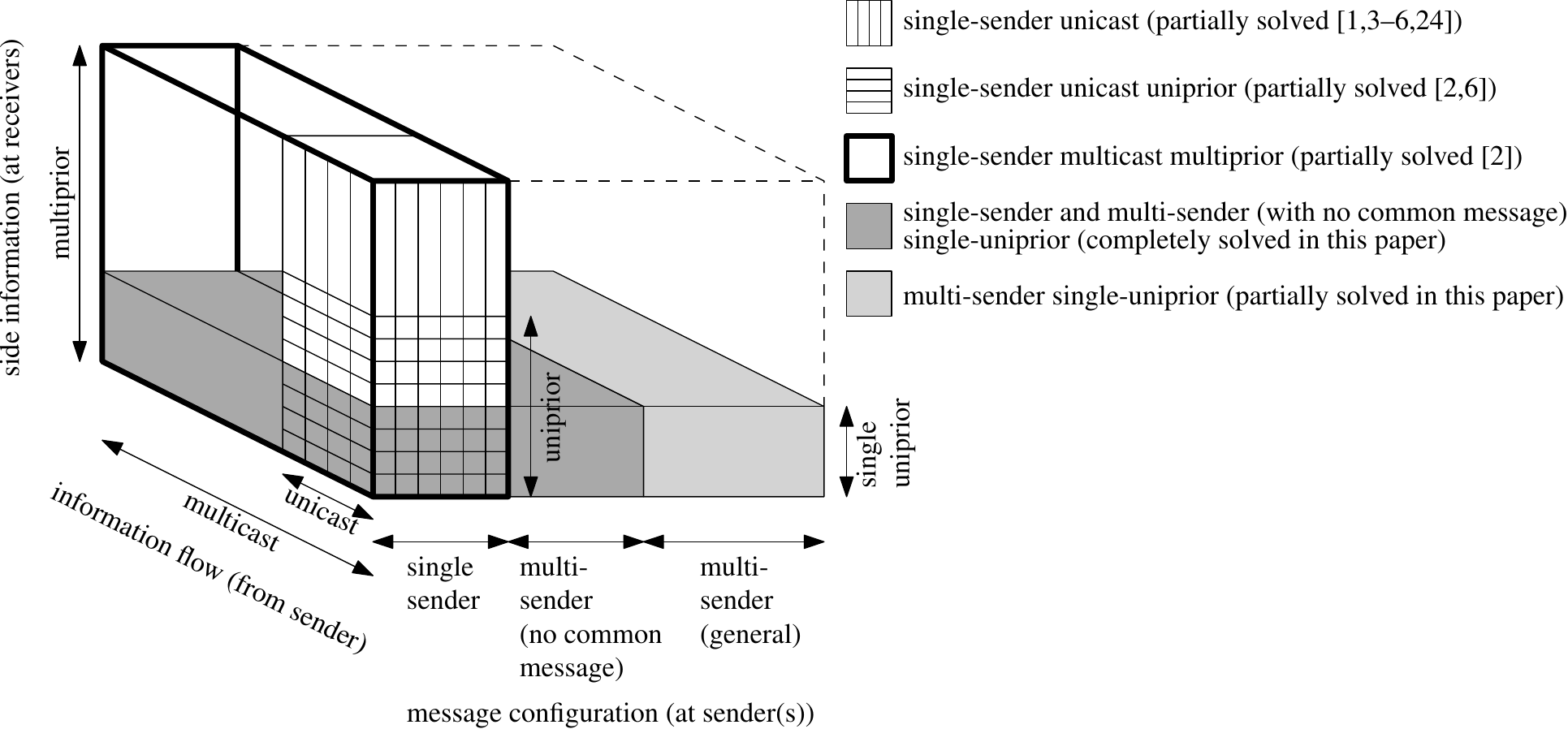}
\caption{Summary of results in this paper and existing results. Note that any unicast instance can be recast as a single-unicast instance. The relative volumes are not indicative of the number of instances in the classes.}
\label{fig:result}
\end{figure*}

\subsection{Related works on multiterminal networks with side information}

In the literature, a problem setup related to the multi-sender index-coding problem is multiterminal networks where a set of clients exchange messages directly through a shared noiseless medium. The aim is to minimize the total transmission \textit{cost} (e.g., total number of transmissions) while satisfying all clients' requirements. This problem can be viewed as a special case of the multi-sender index-coding problem where for each client, there is a sender having the same messages.

Ozgul and Sprintson~\cite{ozgulsprintson11} considered this setup where each client requests \textit{all the messages it does not know}. They used multiple rounds of random linear coding to achieve the optimal transmission cost with high probability.

Hou, Hsu, and Sprintson~\cite{houhsusprintson13} considered the same setup where each client requests \textit{one message (i.e. unicast)}, and it is \textit{selfish} in the sense that it tries to minimize its own transmissions. They proposed a game-theoretic algorithm (via bidding) for transmission, and numerically showed that their algorithm is close to optimal.

Ji, Caire, and Molisch~\cite{jicairemolisch13} also considered this setup where the side information of the clients (which admits a fixed number of bits) can be chosen, but the requests (also with a fixed number of bits) are random variables. They obtained an upper bound and a lower bound (on the optimal \textit{rate}, i.e., total transmitted bits per side-information bits per client)  within a multiplicative gap, by carefully designing the client's side information, such that all requests can be fulfilled by their transmission scheme.

Unlike the setup considered by Ji et al., the setup in this paper does not allow one to design the side information of the receivers. Also, unlike the setups by Ozgul and Sprintson, and Hou et al.,
we consider all possible message requests by the receivers.

\subsection{Our contributions}

\subsubsection{Single-sender}

In this paper, we solve all single-uniprior (including unicast and multicast) index-coding problems, and show that the solution can be found in polynomial time in the worst case. 

More specifically,  we characterize the minimum codelength $\ell^*$ of any single-uniprior index-coding instance in terms of the number of \textit{leaf non-trivial strongly connected components} (SCC) of the corresponding information-flow graph $\mathcal{G}$. 
To this end, we design a \textit{pruning algorithm} that removes arcs from $\mathcal{G}$ to ``destroy'' all leaf non-trivial SCCs, to get a graph through which we derive a lower bound on $\ell^*$. By carefully pruning $\mathcal{G}$, we show that the lower bound is achievable using \textit{linear cyclic codes}. Hence, we incidentally show that linear index codes are optimal for all single-uniprior instances.
This is in stark contrast to the single-unicast instances where non-linear codes can outperform linear codes~\cite{lubertzkystav09}. 

For the rest of this paper, when we say leaf SCCs, we simply mean leaf non-trivial SCCs.

\subsubsection{Multi-sender}

For multi-sender single-uniprior instances, we construct upper and lower bounds on the optimal index codelength based on the information-flow graph $\mathcal{G}$ and the message graph $\mathcal{U}$. For the lower bound, we construct an algorithm that not only prunes $\mathcal{G}$ (using the aforementioned pruning algorithm), but also adds arcs, edges, and vertices, and derive a lower bound on $\tilde{\ell}^*$ through the resultant graph. For the upper bound, we first define a special type of \textit{tree}, and we count the number of such trees that we can fit into the graph. We then propose a coding scheme for each tree and some leaf SCCs to obtain a linear index code. 

We will show that our upper and lower bounds match for a class of multi-sender single-uniprior instances. This class includes as a special case the scenario where no two senders know the same message, i.e., $\mathcal{M}_s \cap \mathcal{M}_t = \emptyset$ for all $s \neq t$.

\subsubsection{Summary of results and techniques}

Table~\ref{table:result} shows the sections in which we derive different bounds and the techniques involved.
 Figure~\ref{fig:result} summarizes our results in relation to other results.

\section{The Single-Sender Single-Uniprior Case}

In this section we focus on the single-sender single-uniprior case.
We will propose a \textit{pruning} algorithm to obtain a lower bound on $\ell^*$, and then construct a linear index code that achieves this lower bound.

For the single-uniprior case, we work on the information-flow graph, denoted by $\mathcal{G} = \{\mathcal{V}, \mathcal{A}\}$, where $\mathcal{V} = \{1,2,\dotsc, n\}$ is the set of vertices, and $\mathcal{A}$ is the set of arcs.
This models an index-coding instance where there are $n$ receivers, and each receiver~$i$ knows $x_i$ a priori and requests $\{x_j: (j \rightarrow i) \in \mathcal{A}\}$.
For the remaining of this paper, we refer to the message $x_i$, known to receiver~$i$, simply as the message \textit{of} vertex $i$. Let $\ell^*(\mathcal{G})$ denote the optimal index codelength for a single-sender uniprior instance represented by $\mathcal{G}$.

\subsection{Lower bound} \label{sec:single-upper}

\subsubsection{Useful lemmas}

We first establish a few lemmas, which we will use to establish our lower bound.

\begin{lemma} \label{lemma:predecessor}
Each receiver~$i \in \mathcal{V}$ must be able to decode the messages of its predecessors in $\mathcal{G}$. 
\end{lemma}

\begin{IEEEproof}
An index code always allows receiver~$i$ to decode $x_j$ if an arc $(j \rightarrow i)$ exists.
In the single-uniprior instance, $x_j$ is all the side information receiver~$j$ has.
Therefore, receiver~$i$ must also be able to decode all the messages that receiver~$j$ is able to decode, i.e., receiver~$i$ can decode all $\mathcal{W}_j = \{x_k: (k \rightarrow j) \in \mathcal{A}\}$. 
Further chaining of this argument proves that receiver~$i$ must be able to decode the messages of the predecessors of vertex~$i$. 
\end{IEEEproof}

\begin{lemma} \label{lemma:predecessor-all}
Every receiver can decode the messages of all predecessors of any leaf vertex, even without utilizing its prior message. 
\end{lemma}

\begin{IEEEproof}
Consider any leaf vertex $i$, its prior message $x_i$ is not requested by any receiver.
So, we can arbitrarily set $x_i = 0$ without affecting the decoding of any receiver, or the codelength.
With this, any receiver (even one without any side information) knows $x_i = 0$, and, by Lemma~\ref{lemma:predecessor}, it must be able to decode all predecessors of the leaf vertex~$i$.
Since the choice of the leaf vertex was arbitrary, we conclude the proof.
\end{IEEEproof}

\subsubsection{A simple lower bound}

Define the set of leaf vertices in $\mathcal{G}$ as $\mathcal{L}(\mathcal{G})$. We have the following from Lemma~\ref{lemma:predecessor-all}:
\begin{lemma} \label{lemma:predecessor-bound}
For any graph $\mathcal{G}$, we have
\begin{equation}
\ell^*(\mathcal{G}) \geq \sum_{i: \text{ $i$ is a predecessor of some vertex in $\mathcal{L}(\mathcal{G})$}} q_i \label{eq:predecessors}
\end{equation}
\end{lemma}
\begin{IEEEproof}
Denote the messages of all predecessors of all leaf vertices by $\boldsymbol{X} \triangleq \{X_i: i$ is a predecessor of some vertex in $\mathcal{L}(\mathcal{G})\}$,\footnote{We use upper-case letters to denote random variables.} and the index code by $\boldsymbol{C}$. 
It follows from Lemma~\ref{lemma:predecessor-all} that any receiver can decode $\boldsymbol{X}$ from $\boldsymbol{C}$, i.e., $H(\boldsymbol{X}|\boldsymbol{C}) = 0$, or $H(\boldsymbol{X},\boldsymbol{C}) = H(\boldsymbol{C})$, where $H(\cdot)$ is the entropy function. It follows that $\ell^*(\mathcal{G}) \geq H(\boldsymbol{C}) = H(\boldsymbol{X},\boldsymbol{C}) \geq H(\boldsymbol{X}) \stackrel{(a)}{=} \sum_{i: i \text{ is a predecessor of some vertex in } \mathcal{L}(\mathcal{G})} H(X_i)  \stackrel{(b)}{=} Q$, where $Q$ is the right-hand side of \eqref{eq:predecessors}. Equality~$(a)$ is derived because the messages are independent, and $(b)$ is derived  because each message bit is uniformly distributed.
\end{IEEEproof}

The lower bound in Lemma~\ref{lemma:predecessor-bound} is not useful when the graph $\mathcal{G}$ contains not many predecessors of leaf vertices. For example, the right-hand side is zero if $\mathcal{G}$ is a cycle. In light of this, we ``process'' the graph to obtain a better lower bound in the following subsection.

\subsubsection{A better lower bound}

We observe the following:
\begin{lemma} \label{lemma:remove-arc-bound}
Let $\mathcal{G} = \{\mathcal{V},\mathcal{A}\}$ and $\mathcal{G}' = \{\mathcal{V}, \mathcal{A}'\}$, where $\mathcal{A}' \subseteq \mathcal{A}$. We have that
\begin{equation}
\ell^*(\mathcal{G}) \geq \ell^*(\mathcal{G}').
\end{equation}
\end{lemma}
\begin{IEEEproof}
Removing arcs in an information-flow graph reduces decoding requirements, while maintaining the prior messages of each receiver. This means any index code for $\mathcal{G}$ is also an index code for $\mathcal{G}'$.
\end{IEEEproof}

Combining Lemmas~\ref{lemma:predecessor-bound} and \ref{lemma:remove-arc-bound} gives the following lower bound:
\begin{lemma} \label{lemma:simple-lower-bound}
Given any $\mathcal{G} = \{\mathcal{V},\mathcal{A}\}$, the optimal index codelength of a single-sender single-uniprior index-coding instance represented by $\mathcal{G}$ is lower bounded as
\begin{equation}
\ell^*(\mathcal{G}) \geq \max_{\mathcal{G}' = \{\mathcal{V}, \mathcal{A}'\} \text{ s.t. }  \mathcal{A}' \subseteq \mathcal{A}} \sum_{i: \text{ $i$ is a predecessor of some vertex in $\mathcal{L}(\mathcal{G}')$}} q_i. \label{eq:simple-lower-bound}
\end{equation}
\end{lemma}

The lower bound in Lemma~\ref{lemma:simple-lower-bound} involves maximizing $\sum q_i$ over all arc-removed subgraphs $\mathcal{G}'$. We now present a way to optimally remove arcs, to attain the right-hand side of \eqref{eq:simple-lower-bound}. To this end, we propose an algorithm to remove certain arcs to obtain a lower bound. Then, in the next section, we will show that this lower bound is indeed achievable, thereby proving that the bound obtained using this algorithm is tight, i.e., that it attains the right-hand side of \eqref{eq:simple-lower-bound}.

We first define the following:
\begin{definition}
A \textit{leaf SCC} is a \textit{non-trivial}\footnote{Recall that a non-trivial SCC has two or more vertices.} SCC that has no outgoing arc (i.e., from a vertex in the SCC to a vertex outside the SCC).
\end{definition}

We now present a lower bound, which is essentially Lemma~\ref{lemma:simple-lower-bound} evaluated with a specific $\mathcal{G}'$:
\begin{theorem}[Single-sender: lower bound] \label{theorem:single-sender-lower}
The optimal index codelength of a single-sender single-uniprior index-coding instance is lower bounded as
\begin{equation}
\ell^*(\mathcal{G}) \geq \sum_{k \in \mathcal{V}} q_k - \sum_{i \in \mathcal{L}(\mathcal{G})} q_i - \sum_{\mathcal{V}_\text{leafSCC($j$)} \in \mathbb{V}} \min_{a \in \mathcal{V}_\text{leafSCC($j$)}} q_a, \label{eq:single-sender-lower-bound}
\end{equation}
where $\mathbb{V} \triangleq \{ \mathcal{V}_\text{leafSCC(1)}, \mathcal{V}_\text{leafSCC(2)}, \dotsc \}$ is the set of all leaf SCCs in $\mathcal{G}$.
\end{theorem}

We defer the proof of Theorem~\ref{theorem:single-sender-lower} to Section~\ref{sec:proof-thm1}, after introducing a pruning algorithm.

\begin{remark}
The lower bound here is more general than the result in our previously-published conference paper~\cite{ongho12}, for which the messages are restricted to be the same size.
\end{remark}

\subsubsection{Proving Theorem~\ref{theorem:single-sender-lower} using a pruning algorithm} \label{sec:proof-thm1}

We start with the following definitions:
\begin{definition}
A vertex is said to be \textit{grounded} if it is either a leaf vertex, or a predecessor of some leaf vertex.
\end{definition}
\begin{definition}
A graph is said to be grounded if every vertex in the graph is grounded.
\end{definition}

\begin{lemma} \label{lemma:no-scc-grounded}
The two statements below are equivalent for any directed graph $\mathcal{G}$:
\begin{enumerate}
\item $\mathcal{G}$ has no leaf SCC.
\item $\mathcal{G}$ is grounded. 
\end{enumerate}
\end{lemma}

\begin{IEEEproof}
See Appendix~\ref{appendix:grounded}.
\end{IEEEproof}


We now propose an arc-removing algorithm, called the \textit{pruning algorithm}, which we will use to prove Theorem~\ref{theorem:single-sender-lower}. 
\begin{algorithm}[h]
\SetKwInOut{Input}{input}
\SetKwInOut{Output}{output}

\Input{A directed graph $\mathcal{G} = \{\mathcal{V}, \mathcal{A}\}$ with vertex weights $\{q_i\}_{i \in \mathcal{V}}$}
\Output{A grounded directed graph}
\BlankLine
\ForEach{Leaf SCC, say \textnormal{$\mathcal{G}_\text{leafSCC} = \{ \mathcal{V}_\text{leafSCC}, \mathcal{A}_\text{leafSCC}\} \subseteq \mathcal{G}$}}{
Arbitrarily select a vertex with the least weight, i.e., any $i \in \argmin_{a \in \mathcal{V}_\text{leafSCC}} q_a$\;
Remove all outgoing arcs from vertex $i$, i.e., all $(i \rightarrow k) \in \mathcal{A}$\;
}
\caption{The Pruning Algorithm}
\label{algo:pruning}
\end{algorithm}

Executing the pruning algorithm gives the following result:
\begin{lemma} \label{lemma:pruning}
Consider a graph $\mathcal{G} = \{\mathcal{V}, \mathcal{A}\}$. Let the resultant graph after running the pruning algorithm be $\mathcal{G}^\text{p} = \{\mathcal{V}^\text{p}, \mathcal{A}^\text{p}\}$. We have
\begin{align}
&\mathcal{V}^\text{p} = \mathcal{V}, \label{eq:pruning001} \\
&\mathcal{A}^\text{p} \subseteq \mathcal{A}, \label{eq:pruning002} \\
&\sum_{i: \text{ $i$ is a predecessor of some vertex in $\mathcal{L}(\mathcal{G}^\text{p})$}} q_i \notag \\
&\quad\quad =  \sum_{k \in \mathcal{V}} q_k - \sum_{i \in \mathcal{L}(\mathcal{G})} q_i - \sum_{\mathcal{V}_\text{leafSCC($j$)} \in \mathbb{V}} \min_{a \in \mathcal{V}_\text{leafSCC($j$)}} q_a. \label{eq:pruning003}
\end{align}
\end{lemma}

\begin{IEEEproof}
Since the algorithm only removes arcs, we have \eqref{eq:pruning001} and \eqref{eq:pruning002}.

We now show that $\mathcal{G}^\text{p}$ has no leaf SCC. Consider a leaf SCC in $\mathcal{G}$ for which vertex $i$ is selected and all its outgoing arcs are removed. By definition, every vertex in the SCC has a path to $i$. After the \textbf{foreach} iteration, $i$ is made a leaf vertex, and  all vertices in the leaf SCC are therefore grounded. Also, as all SCCs are vertex disjoint, this iteration
\textit{destroys} one leaf SCC and does not create any new leaf SCC. When the algorithm terminates, $\mathcal{G}^\text{p}$ has no leaf SCC, and it follows from Lemma~\ref{lemma:no-scc-grounded} that $\mathcal{G}^\text{p}$ is grounded.


Since $\mathcal{G}^\text{p}$ is grounded, each non-leaf vertex is a predecessor of some leaf vertex.
This means
\begin{align}
\sum_{i: \text{ $i$ is a predecessor of some vertex in $\mathcal{L}(\mathcal{G}^\text{p})$}} q_i &= \sum_{i \in \mathcal{V} \setminus \mathcal{L}(\mathcal{G}^\text{p})} q_i \notag \\
& = \sum_{k \in \mathcal{V}} q_k - \sum_{i \in \mathcal{L}(\mathcal{G}^\text{p})}q_i. \label{eq:lower-bound-g'}
\end{align}


Recall that $\mathbb{V} \triangleq \{\mathcal{V}_\text{leafSCC(1)}, \mathcal{V}_\text{leafSCC(2)}, \dotsc \}$ is the set of all leaf SCCs in $\mathcal{G}$.
These leaf SCCs are vertex-disjoint by definition. In Algorithm~\ref{algo:pruning}, one vertex (one with the least weight) in each leaf SCC in $\mathcal{G}$ is made a leaf vertex. So, 

\begin{equation}
\sum_{i \in \mathcal{L}(\mathcal{G}^\text{p})}q_i  = \sum_{i \in \mathcal{L}(\mathcal{G})} q_i + \sum_{\mathcal{V}_\text{leafSCC($j$)} \in \mathbb{V}} \min_{a \in \mathcal{V}_\text{leafSCC($j$)}} q_a. \label{eq:all-non-leaf}
\end{equation}

Substituting \eqref{eq:all-non-leaf} into  \eqref{eq:lower-bound-g'}, we have \eqref{eq:pruning003}.
\end{IEEEproof}


\begin{IEEEproof}[Proof of Theorem~\ref{theorem:single-sender-lower}]
Combining Lemmas~\ref{lemma:simple-lower-bound} and~\ref{lemma:pruning}, we have Theorem~\ref{theorem:single-sender-lower}.
\end{IEEEproof}

\begin{remark}
 \textit{(Complexity of the pruning algorithm)} 
All SCCs in a graph can be found in linear time in $|\mathcal{V}|+|\mathcal{A}|$ (see, for example, the Kosaraju-Sharir algorithm~\cite{sharir81} or Tarjan's algorithm~\cite{tarjan72}). To check if an SCC is a leaf SCC, we check whether all out-neighbors of each vertex in the SCC is in the same SCC. This takes at most $|\mathcal{A}||\mathcal{V}|$ checks. Finding the vertex with the minimum message length in a leaf SCC can be done by a sorting algorithm with a worst-case complexity of $|\mathcal{V}|\log |\mathcal{V}|$. So, the pruning algorithm runs in $O(|\mathcal{V}|^3)$ in the worst case.
\end{remark}

The main idea behind our lower bound here is that removing some requests by a receiver (i.e., reducing the sets $\{\mathcal{W}_r\}$) cannot increase the optimal index codelength. This corresponds to removing arcs in the information-flow graph. A similar---but different---concept was used to derive a lower bound for the unicast index-coding instances (which are represented by side-information graphs), namely, the maximum-acylic-induced-subgraph (MAIS) lower bound~\cite{baryossefbirk11,neelytehranizhang12}. The main idea behind the MAIS bound is that removing messages (i.e., reducing $\mathcal{M}$ and adjusting $\{\mathcal{K}_r,\mathcal{W}_r\}$ accordingly) from the system cannot increase the optimal index codelength. This corresponds to a vertex-induced subgraph.
While the MAIS lower bound is in general not tight for single-sender \textit{unicast} index-coding instances~\cite{blasiakkleinberglubertzky13}, we will see that our proposed lower bound based on removing arcs is tight for all single-sender \textit{uniprior} index-coding instances.

\subsection{Upper bound (achievability): cyclic codes} \label{sec:single-lower}

In this subsection, we construct linear index codes that achieve the lower bound given in Theorem~\ref{theorem:single-sender-lower}. Although the lower bound is calculated based on the pruned graph $\mathcal{G}^\text{p}$, for achievability, we need to design index codes for the original graph $\mathcal{G}$. It turns out that by coding on leaf SCCs in $\mathcal{G}$, we can construct index codes that attain the lower bound. 

It is easier to discuss the lower bound, i.e., the right-hand side of (\ref{eq:single-sender-lower-bound}), in terms of a trivial index code and ``savings''. To see this,  we compare it with a trivial index code that simply transmits $\sum\limits_{k \in \mathcal{V}} q_k - \sum\limits_{i \in \mathcal{L}(\mathcal{G})} q_i$ bits, which are the uncoded messages of the non-leaf vertices (since, by definition, the messages of the leaf vertices are not requested by any receiver). 
We will further show that an extra savings of $\sum\limits_{\mathcal{V}_\text{leafSCC($j$)} \in \mathbb{V}} \min\limits_{a \in \mathcal{V}_\text{leafSCC($j$)}} q_a$ bits is possible---a consequence of the leaf SCCs.
This corresponds to a savings, over the trivial index code, of $\min\limits_{a \in \mathcal{V}_\text{leafSCC}} q_a$ bits for each leaf SCC $\mathcal{V}_{\text{leafSCC}} \in \mathbb{V}$.

We will use \textit{cyclic codes}, defined below, to realize the required savings:
\begin{definition}[Cyclic codes]
Consider $n$ $q$-bit messages, $\{x_1, x_2, \dotsc, x_n\}$. A cyclic code is a length $q(n-1)$-bit code, constructed as follows:
\begin{equation*}
x_1 \oplus x_2, \;\;
x_2 \oplus x_3,  \;\;
\dotsc,\;\;
x_{n-1} \oplus x_n,
\end{equation*}
where $\oplus$ bit-wise XOR. 
\end{definition}

\begin{theorem}[Single-sender: upper bound] \label{theorem:single-sender-upper}
The optimal index codelength of a single-sender single-uniprior index-coding instance is upper bounded as
\begin{equation}
\ell^*(\mathcal{G}) \leq \sum_{k \in \mathcal{V}} q_k - \sum_{i \in \mathcal{L}(\mathcal{G})} q_i - \sum_{\mathcal{V}_\text{leafSCC($j$)} \in \mathbb{V}} \min_{a \in \mathcal{V}_\text{leafSCC($j$)}} q_a.
\end{equation}
\end{theorem}

\begin{IEEEproof}[Proof of Theorem~\ref{theorem:single-sender-upper}]
For each non-leaf vertex $i$ that does not belong to any leaf SCC, we send $x_i$ uncoded. 

For vertices in leaf SCCs, we construct the following code:
Consider a leaf SCC in $\mathcal{G}$, and let its vertices be $\mathcal{V}_\text{leafSCC} = \{1, 2, \dotsc, v\}$.
For now, let us first assume all messages $\{x_{i}\}_{i=1}^v$ in the leaf SCC are of equal length, say $q$ bits.
Then construct the following cyclic code:
\begin{equation} \label{eq:cyclic}
    x_{1} \oplus x_{2}, \;\;
    x_{2} \oplus x_{3},\;\; \dotsc, \;\;
    x_{v-1} \oplus x_{v}.
\end{equation}

Note that for any $i \in \mathcal{V}_\text{leafSCC}$, receiver~$i$
\begin{itemize}
\item can decode all messages in $\mathcal{V}_\text{leafSCC}$ from \eqref{eq:cyclic} and its side information $x_i$,
\item can also decode the messages of all non-leaf vertices not in any leaf SCC (as these messages are sent uncoded), and
\item does not request any messages in other leaf SCCs (due to the definition of leaf SCC).
\end{itemize}

Repeating the above cyclic code for each leaf SCC, we satisfy the decoding requirements for all receivers in all leaf SCCs.
Also, any receiver not in any leaf SCC can also decode its requested messages, as these messages are sent uncoded. Note that, by definition, messages in any leaf SCC are not requested by any receiver outside the leaf SCC.

To generalize the above to non-equal-length messages of $q_i$ bits for each vertex~$i$, simply form \eqref{eq:cyclic} by substituting each $x_i$ with its first $q_\text{min}$ bits, where $q_\text{min}$ is the shortest message length in the leaf SCC, i.e., $q_\text{min} = \min\limits_{1 \leq a \leq v} q_a$. 
Then the total number of coded bits in \eqref{eq:cyclic} is $(v-1)q_\text{min}$ bits.
We send the remaining bits $\sum_{i=1}^vq_{i} - v q_\text{min}$ uncoded.
Hence, the total savings per leaf SCC is $q_\text{min}$.

Repeating this for each leaf SCC, we save a total of $\sum\limits_{\mathcal{V}_\text{leafSCC($j$)} \in \mathbb{V}} \min\limits_{a \in \mathcal{V}_\text{leafSCC($j$)}} q_a$ bits for the entire graph $\mathcal{G}$.
\end{IEEEproof}

\begin{remark}
The cyclic code \eqref{eq:cyclic} used in the proof of Theorem~\ref{theorem:single-sender-upper} can be derived from the partial-clique-cover scheme~\cite{birkkol2006} (originally proposed for unicast index coding). Consider \textit{(i)} $v$ messages $\{x_i\}_{i=1}^v=\mathcal{M}$ each taking values from a finite field $\mathbb{F}$; \textit{(ii)} a set of receivers $\{1,2,\dotsc,n\}$, where each receiver $i \in \{1,2,\dotsc,n\}$ knows a priori at least $k$ messages in $\mathcal{M}$, i.e., $|\mathcal{K}_i \cap \mathcal{M}| \geq k$. The partial-clique-cover scheme gives a codeword based on minimum-distance-separable codes with the following properties: \textit{(a)} it consists of $(v-k)$ finite-field symbols; \textit{(b)} it simultaneously lets each receiver decode all messages in $\mathcal{M}$. For a leaf SCC in a single-uniprior index-coding problem, we have $k=1$. By setting $\mathbb{F}=\{0,1\}$, and appropriately constructing a minimum-distance-separable code, one can obtain \eqref{eq:cyclic} for $q=1$ (a codeword of $(v-1)$ bits), which allows all receivers in the SCC to decode all messages in the SCC. The multiple-bit version of \eqref{eq:cyclic} can be obtained by concatenating $q>1$ copies of the binary codewords.
\end{remark}

\subsection{The optimal index codelength} 

Combining Theorems~\ref{theorem:single-sender-lower} and \ref{theorem:single-sender-upper}, we have the following:
\begin{theorem} \label{theorem:single-sender}
The optimal index codelength of a single-sender single-uniprior index-coding instance is
\begin{equation}
\ell^*(\mathcal{G}) = \sum_{k \in \mathcal{V}} q_k - \sum_{i \in \mathcal{L}(\mathcal{G})} q_i - \sum_{\mathcal{V}_\text{leafSCC($j$)} \in \mathbb{V}} \min_{a \in \mathcal{V}_\text{leafSCC($j$)}} q_a.
\end{equation}
Furthermore, linear codes can achieve the optimal index codelength.
\end{theorem}

\ifx\doublecolumn\undefined
\begin{figure}[t]
\centering
\resizebox{18ex}{!}{%
\begin{tikzpicture}
\def\S{0.4}
\def\D{1.5}
\def\A{10}
\graph {[grow right=\D, branch down=\D, nodes={draw,circle},edge={>=latex}]
1 ->[bend right=\A] {2, 3[not target]} <- {4, 5[not source]},
2 ->[bend right=\A] 1,
3 ->[bend right=\A] 1,
2 ->[bend right=\A] 3,
3 ->[bend right=\A] 2,
4 -> 5,
};
\node at ({-\S},\S) {\scriptsize 1};
\node at ({\D+\S},\S) {\scriptsize 2};
\node at ({2*\D+\S},\S) {\scriptsize 2};
\node at ({\D+\S},{-\D-\S}) {\scriptsize 2};
\node at ({2*\D+\S},{-\D-\S}) {\scriptsize 2};
\end{tikzpicture}
}%
\caption{An example of a single-uniprior index-coding instance}
\label{fig:example-2}
\end{figure}
\else
\begin{figure}[t]
\centering
\includegraphics[width=5cm]{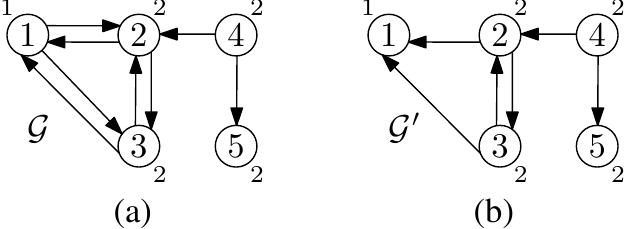}
\caption{An example of a single-uniprior index-coding instance: (a) Graph $\mathcal{G}$ is an information-flow graph representing the problem instance. (b) Graph $\mathcal{G}'$ is the resultant graph after running the pruning algorithm on $\mathcal{G}$}
\label{fig:example-2}
\end{figure}
\fi

\begin{example} \label{ex:single-sender}
We illustrate how to obtain the optimal codelength and an optimal code for an index-coding instance with one single-bit message (i.e., $x_1 = x_1[1]$), four two-bit messages (i.e., $x_i = [x_i[1], x_i[2]]$ for $i \in \{2,3,4,5\}$), and five receivers. Each receiver~$i$ knows $\mathcal{K}_i = \{x_i\}$, and the receivers each request the following messages: $\mathcal{W}_1 = \{x_2,x_3\}, \mathcal{W}_2 = \{x_1, x_3, x_4\}, \mathcal{W}_3 = \{x_1,x_2\}, \mathcal{W}_4 = \emptyset, \mathcal{W}_5 = \{x_4\}$.
This problem can be represented by a graph~$\mathcal{G}$ depicted in Figure~\ref{fig:example-2}. It contains one leaf SCC with vertices $\mathcal{V}_\text{leafSCC} = \{1,2,3\}$, in which message $x_1$ has the shortest length of one bit. There is only one leaf vertex in $\mathcal{G}$, which is node 5.   
Theorem~\ref{theorem:single-sender-lower} gives a lower bound $\ell^*(\mathcal{G}) \geq 9 - 2 - 1 = 6$.
To construct an optimal index code, we form a two-bit cyclic code $( x_1[1] \oplus x_2[1], x_2[1] \oplus x_3[1])$ for the leaf SCC in $\mathcal{G}$, and four uncoded bits for the remaining messages in the leaf SCC and the messages of the other the non-leaf vertices $(x_2[2], x_3[2], x_4[1], x_4[2])$. 
Here, $\ell^*(\mathcal{G}) = 6$.
\end{example}

For the special cases where all messages are binary, Theorem~\ref{theorem:single-sender} simplifies to the following:
\begin{corollary} \label{corollary:single-sender-binary}
The optimal index codelength of a single-sender single-uniprior index-coding instance with \textit{binary messages} is
\begin{equation}
    \ell^*(\mathcal{G}) = n - |\mathcal{L}(\mathcal{G})|- \total(\mathcal{G}), \label{eqn:single-sender-binary}
\end{equation}
where $|\mathcal{L}(\mathcal{G})|$ is the number of leaf vertices in $\mathcal{G}$ and $\total(\mathcal{G})$ is the total number of leaf SCCs in $\mathcal{G}$.
\end{corollary}


\section{The Multi-Sender Case: A Lower Bound} \label{sec:multiple-lower}

In this section, we will modify the pruning algorithm to include information from $\mathcal{U}$, and further propose an \textit{appending} step to modify the graph $\mathcal{G}$, based on which we derive a lower bound for the multi-sender case. 

For simplicity and clarity, we will only consider the case where all messages $x_i$ are binary.
This means all vertices in the information-flow graph $\mathcal{G}$ all have weight one. 
One can extend the ideas developed in this section to the more general case of unequal message sizes.

\subsection{The single-sender bound may be loose for multi-sender}

Denote a multi-sender instance by $(\mathcal{G}, \mathcal{U})$. 
The optimal codelength $\tilde{\ell}^*(\mathcal{G},\mathcal{U})$ clearly is lower bounded by $\ell^*(\mathcal{G})$ for the single-sender case, because a single sender who knows all the messages can also send any multi-sender index code. We also know from Lemma~\ref{lemma:predecessor-bound} that, for a graph $\mathcal{G}$ with binary messages, $\ell^*(\mathcal{G})$ is lower bounded by the number of vertices that are each a predecessor of some leaf vertex, denoted by $\predecessor(\mathcal{G})$. This gives
\begin{equation}
\tilde{\ell}^*(\mathcal{G},\mathcal{U}) \geq \ell^*(\mathcal{G}) \geq \predecessor(\mathcal{G}). \label{eq:compare-single-bit-multi-bit}
\end{equation}

For the single-sender problem, we used the pruning algorithm to prune all leaf SCCs to get a grounded graph $\mathcal{G}^\text{p}$. 
As pruning does not increase the optimal codelength (see Lemma~\ref{lemma:remove-arc-bound}), we have the following single-sender bound for multi-sender instances:
\begin{equation}
\tilde{\ell}^*(\mathcal{G},\mathcal{U}) \geq \ell^*(\mathcal{G}) \geq \ell^*(\mathcal{G}^\text{p}) \geq \predecessor(\mathcal{G}^\text{p}) = V_\text{out}(\mathcal{G}^\text{p}),  \label{eq:single-sender-lower-bounds-multi}
\end{equation}
where $V_\text{out}(\mathcal{G})$ denotes the total number of non-leaf vertices in $\mathcal{G}$. The equality above is derived as $\mathcal{G}^\text{p}$ is grounded.


However, the single-sender lower bound may be loose due to the sender constraints specified by the undirected graph $\mathcal{U}$ (see its definition in Section \ref{section:multi-sender-ext}). 
Consider Example~\ref{ex:single-sender} again, and suppose that there are five senders, where each sender $i \in \{1,2,\dotsc,5\}$ has only $x_i$.
Under this constraint, it is not possible to transmit $x_1[1] \oplus x_2[1]$, as the messages belong to different senders. 
The minimum codelength for this case is seven bits---sending $[x_1,x_2,x_3,x_4]$ uncoded. 
On the other hand, if some sender has $\{x_1,x_2\}$ and another has $\{x_2,x_3\}$, then the six-bit lower bound is tight using the coding scheme mentioned in the example.

\subsection{A tighter bound by appending}





We will propose a new algorithm that gives a better lower bound.
 Our new algorithm produces a resultant graph, say $\mathcal{G}^\dagger$, with a possibly higher number of $\predecessor(\mathcal{G}^\dagger) > \predecessor(\mathcal{G}^\text{p})$, where $\mathcal{G}^\text{p}$ is the resultant graph after running Algorithm~\ref{algo:pruning}.
To obtain the desired $\mathcal{G}^\dagger$, we propose a new technique that \textit{appends} some leaf SCCs, by adding an outgoing arc from each of these leaf SCC to another vertex or a newly introduced grounded vertex. 
In the new algorithm, we iterate on the leaf SCCs (either pruning or appending) until we get a grounded information-flow graph $\mathcal{G}^\dagger$.\footnote{A lower bound based on a non-grounded resultant graph is strictly suboptimal, because the graph must contain at least one leaf SCC, and by pruning the leaf SCC, we can increase $\predecessor(\cdot)$.}
From the resultant $\mathcal{G}^\dagger$, we get
\begin{equation}
    \tilde{\ell}^*(\mathcal{G}^\dagger,\mathcal{U}^\dagger) \geq \ell^*(\mathcal{G}^\dagger) \geq \predecessor(\mathcal{G}^\dagger) = V_\text{out}(\mathcal{G}^\dagger), \label{eqn:prune-append-lower-bound}
\end{equation}
for all $\mathcal{U}^\dagger$, where the inequalities follow from \eqref{eq:compare-single-bit-multi-bit}, and the equality is derived because $\mathcal{G}^\dagger$ is grounded. 

To link this lower bound to the original graph $(\mathcal{G},\mathcal{U})$, we need to make sure that the optimal index codelength cannot increase after each pruning/appending step.
If this is satisfied, we have the lower bound
\begin{subequations}
\begin{align}
\tilde{\ell}^*(\mathcal{G},\mathcal{U}) & \geq \tilde{\ell}^*(\mathcal{G}^\dagger,\mathcal{U}^\dagger) \label{eq:index-code-cannot-increase}\\
&  \geq V_\text{out}(\mathcal{G}^\dagger). \label{eq:proposed-new-lower-bound}
\end{align}
\end{subequations}
We aim to maximize the number of non-leaf vertices in the resultant graph (which is grounded).

We will see that pruning reduces the number of non-leaf vertices by one, while appending does not change the number of non-leaf vertices. So, potentially $V_\text{out}(\mathcal{G}^\dagger) > V_\text{out}(\mathcal{G}^\text{p})$, which means the lower bound \eqref{eq:proposed-new-lower-bound} is potentially tighter than \eqref{eq:single-sender-lower-bounds-multi}. 
We have seen in the single-sender case that \eqref{eq:index-code-cannot-increase} is always true if we prune all leaf SCCs. 
However, appending in general increases decoding requirements, which may cause the optimal index codelength to increase.
The main challenge here is to append certain types of leaf SCCs such that the optimal index codelength cannot increase, i.e.,  \eqref{eq:index-code-cannot-increase} holds. 
To this end, we will classify leaf SCCs based on $\mathcal{U}$.

\subsection{Classifying leaf SCCs to reflect the sender setting} \label{sec:classification}


We now give some intuition on how we use $\mathcal{U}$ to classify the leaf SCCs. 
Recall that $\mathcal{U}$ is an undirected graph where an edge exists between vertices $i$ and $j$, if and only if some sender has both $x_i$ and $x_j$.
Connectivity in $\mathcal{U}$ restricts the construction of index codes, which allows us to append some leaf SCCs. For example, 
  if there is no path between two vertices in $\mathcal{U}$, say vertices 1 and 2, 
then it turns out that
  we can partition any index codeword into two sub-codewords, where one sub-codeword is a function of only messages from a set, say $\mathcal{M}' \subset \mathcal{M}$, containing $x_1$, and the other sub-codeword a function of the only messages from the set $\mathcal{M} \setminus \mathcal{M}'$, which contains $x_2$. This, we will show, leads to the result that any receiver can decode all messages in a leaf SCC that contains two vertices not connected in $\mathcal{U}$. As a consequence, we can append this type of leaf SCC, while satisfying \eqref{eq:index-code-cannot-increase}, by adding an arc from the leaf SCC to a dummy vertex.

We will now formally classify different types of leaf SCCs based on their connectivity in $\mathcal{U}$.
We first define \textit{neighboring vertices} in $\mathcal{U}$. For a vertex set $\S\subseteq\mathcal{V}$, we say that a vertex $i \notin \S$ is a \emph{neighbor} of $\S$ if and only if there is an edge $(i,v)$ in $\mathcal{U}$ between $i$ and some $v \in \S$. 
\begin{enumerate}
\item \textbf{Message-connected leaf SCC:}
A leaf SCC in $\mathcal{G}$ is said to be message-connected if and only if there always exists a path\footnote{Recall that we use \textit{path} to denote an undirected path (of edges) in $\mathcal{U}$, or a \textit{directed path} (of arcs) in $\mathcal{G}$.}  in $\mathcal{U}$ between any two vertices in the SCC, where the path consists of vertices only in the SCC.
\item \textbf{Message-disconnected leaf SCC:}
A leaf SCC is message-disconnected if and only if there are two vertices in the SCC with no path in $\mathcal{U}$ between them (even if the path can contain vertices outside the SCC).
\item \textbf{Semi-message-connected leaf SCC}
A leaf SCC that is neither message-connected nor message-disconnected is semi-message-connected, referred also as semi leaf SCC for short. Here, we can always find a vertex pair, where all paths between them must contain some vertex outside the SCC. We further classify semi leaf SCCs:
\begin{enumerate}
\item \textbf{Degenerated leaf SCC:}
A semi leaf SCC, with vertex set $\S$, is said to be \textit{degenerated} if and only if we can find two vertex sets
\begin{itemize}
\item $\Sinside \subset \mathcal{V}_\text{S}$, which is inside the leaf SCC,  and
\item $\Soutside \subseteq \mathcal{V} \setminus \mathcal{V}_\text{S}$, which is outside the leaf SCC,
\end{itemize}
such that
\begin{itemize}
\item  there is no edge in $\mathcal{U}$ between $\Sinside$ and $\mathcal{V}_\text{S} \setminus \Sinside$ (the complement of $\Sinside$ in the leaf SCC), 
\item  there is at most one non-leaf vertex in $\Soutside$, and
\item  every neighbor of $\Sinside$ is either in $\Soutside$, or a predecessor (with respect to the directed graph $\mathcal{G}$) of some vertex in $\Soutside$.
\end{itemize}
\item \textbf{Non-degenerated leaf SCCs:} A semi leaf SCC that is not degenerated is said to be non-degenerated.
\end{enumerate}
\end{enumerate}

\ifx\doublecolumn\undefined
\begin{figure}[t]
\centering
\includegraphics[width=11cm]{mwrc-it-04}
\caption{Classification of leaf SCCs: By definition, leaf SCCs are determined by $\mathcal{G}$ (where arcs are drawn with arrows), but their various types are determined also in accordance with $\mathcal{U}$ (where edges are drawn with solid lines).
This graph illustrates concurrently three leaf SCC types: (i) message-connected, where there is a path between any two vertices through only vertices in the SCC; (ii) message-disconnected, containing at least two vertices that cannot be connected by any path; and (iii) semi-message-connected, where some vertices must be connected by a path with vertices outside the SCC. The semi leaf SCC here (with vertex set $\S$) is degenerated because we can find two vertex sets, $\Sinside$ in the leaf SCC and $\Soutside$ outside the SCC, such that (a) there is no edge between $\Sinside$ and $\mathcal{V}_\text{S} \setminus \Sinside$, (b) $\Soutside$ contains at most one non-leaf vertex, and (c) all neighbors of $\Sinside$  are predecessors of $\Soutside$.}
\label{fig-1}
\end{figure}
\else
\begin{figure}[t]
\centering
\includegraphics[width=\linewidth]{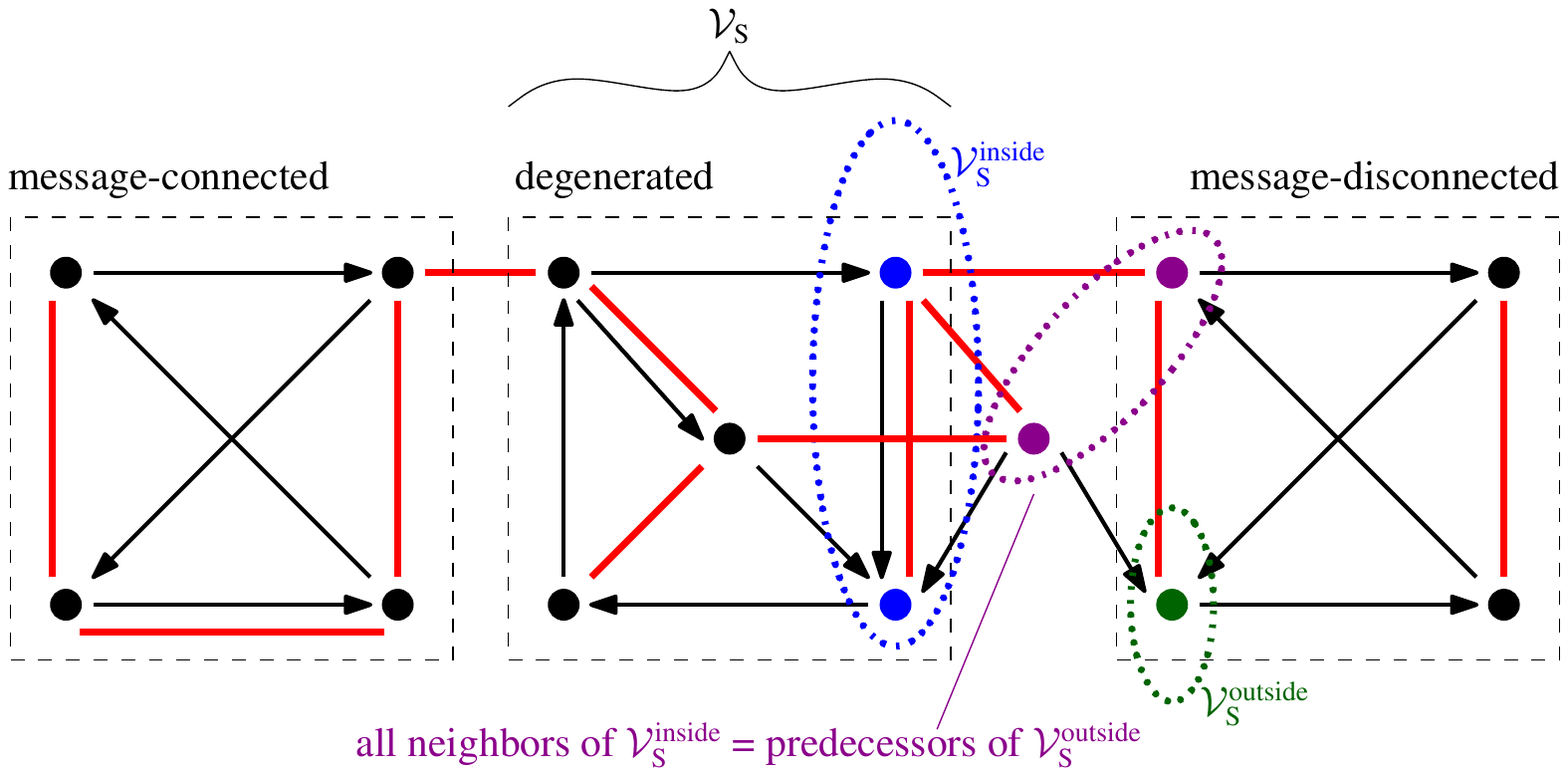}
\caption{Classification of leaf SCCs: By definition, leaf SCCs are determined by $\mathcal{G}$ (where arcs are drawn with arrows), but their various types are determined also in accordance with $\mathcal{U}$ (where edges are drawn with solid lines).
This graph illustrates concurrently three leaf SCC types: (i) message-connected, where there is a path between any two vertices through only vertices in the SCC; (ii) message-disconnected, containing at least two vertices that cannot be connected by any path; and (iii) semi-message-connected, where some vertices must be connected by a path with vertices outside the SCC. The semi leaf SCC here (with vertex set $\S$) is degenerated because we can find two vertex sets, $\Sinside$ in the leaf SCC and $\Soutside$ outside the SCC, such that (a) there is no edge between $\Sinside$ and $\mathcal{V}_\text{S} \setminus \Sinside$, (b) $\Soutside$ contains at most one non-leaf vertex, and (c) all neighbors of $\Sinside$  are predecessors of $\Soutside$. }
\label{fig-1}
\end{figure}
\fi

Any leaf SCC must belong to one of these four types. Figure~\ref{fig-1} shows examples of three types of leaf SCCs.
As cryptic as the definition of a degenerated leaf SCC may seem, it has been carefully crafted to allow us to append it.

\subsection{Which leaf SCCs to append and which to prune?}

Recall that our lower bound is given by $\tilde{\ell}^*(\mathcal{G},\mathcal{U}) \geq V_\text{out}(\mathcal{G}^\dagger)$, and we want the resultant grounded information-flow graph $\mathcal{G}^\dagger$ to have as many non-leaf vertices as possible. 
As mentioned before, we would prefer appending to pruning, as the latter always reduces the number of non-leaf vertices, but the former does not.
However, for message-connected and non-degenerated leaf SCCs, we are unable to show that appending will not increase the optimal index codelength, as required by \eqref{eq:index-code-cannot-increase}. In fact, appending a message-connected leaf SCC can strictly increase the optimal index codelength, as shown by the following example:

\begin{example}
Consider a single-uniprior index-coding instance, whose information-flow graph is given by
\vspace{2ex}
\begin{center}
\resizebox{1.9cm}{!}{
\begin{tikzpicture}
\def\S{0.4}
\def\D{1.5}
\def\A{10}
\graph {[grow right=\D, branch down=\D, nodes={draw,circle},edge={>=latex}]
1 ->[bend right] 2,
2 ->[bend right] 1,
};
\end{tikzpicture}
}
\end{center}
\vspace{1ex}
and a sender has both the messages. Here, the graph itself is a message-connected leaf SCC. The optimal codelength is one, attained by $x_1 \oplus x_2$. The single-sender lower bound $\tilde{\ell}^*(\mathcal{G},\mathcal{U})  \geq 1$ is tight. If we append (we will formally define the appending step in the next section) this leaf SCC, by creating a dummy vertex 3, and an arc $(2 \rightarrow 3)$, the resultant grounded graph $\mathcal{G}^\dagger$ is
\vspace{2ex}
\begin{center}
\resizebox{3.1cm}{!}{
\begin{tikzpicture}
\def\S{0.4}
\def\D{1.5}
\def\A{10}
\graph {[grow right=\D, branch down=\D, nodes={draw,circle},edge={>=latex}]
1 ->[bend right] 2 -> 3,
2 ->[bend right] 1,
};
\end{tikzpicture}
}
\end{center}
\vspace{1ex}
and we get $\tilde{\ell}^*(\mathcal{G}^\dagger,\mathcal{U}^\dagger) \geq \predecessor(\mathcal{G}^\dagger) = 2$. For this example, \eqref{eq:index-code-cannot-increase} is not true, and $\predecessor(\mathcal{G}^\dagger)$ is not a lower bound on $\tilde{\ell}^*(\mathcal{G},\mathcal{U})$.
\end{example}

So, we will
\begin{itemize}
\item prune message-connected and non-degenerated leaf SCCs, and
\item append message-disconnected and degenerated leaf SCCs.
\end{itemize}

In the following three subsections, we will prove that appending message-disconnected and degenerated leaf SCCs indeed guarantees \eqref{eq:index-code-cannot-increase}. We will also derive the number of non-leaf vertices after each pruning/appending step, and show that the number of leaf SCCs that remain in the graph cannot increase (so that our algorithm always terminates).

We will use the following notation. Let $\mathcal{G}$ and $\mathcal{U}$ be the information-flow graph and the message graph (respectively) before a pruning/appending step, and $\mathcal{G}'$ and $\mathcal{U}'$ be the respective graphs after the step.

\subsubsection{Appending a message-disconnected leaf SCC} \label{sec:prop1}


\begin{definition}[Appending a message-disconnected leaf SCC]
To append a message-disconnected leaf SCC (with a vertex set $\S$), we add
\begin{itemize}
\item a dummy vertex $n+1$ in both $\mathcal{G}$ and $\mathcal{U}$ where $x_{n+1} = 0$, which is known to all receivers and is hence never transmitted by the sender,  and
\item a dummy arc in $\mathcal{G}$ from an arbitrarily chosen vertex in the leaf SCC to the dummy vertex, i.e., choose some $v \in \S$ and add $(v \rightarrow n+1)$ in $\mathcal{G}$.
\end{itemize}
\end{definition}

\begin{proposition} \label{prop:disconnected}
After appending a message-disconnected leaf SCC, we have
\begin{align}
\total(\mathcal{G}') &= \total(\mathcal{G}) - 1, \label{eq:decrease-n-total-1}\\
\tilde{\ell}^*(\mathcal{G}',\mathcal{U}') &= \tilde{\ell}^*(\mathcal{G},\mathcal{U}), \label{eq:decrease-opt-length-1}\\
V_\text{out}(\mathcal{G}') &= V_\text{out}(\mathcal{G}). \label{eq:equal-v-out-1}
\end{align}
\end{proposition}

\begin{IEEEproof}
See Appendix~\ref{appendex:proof-prop-disconnected}.
\end{IEEEproof}

\subsubsection{Appending a degenerated leaf SCC} \label{sec:prop2}

Recall that a semi leaf SCC (with vertex set $\S$) is degenerated if and only if we can find two vertex sets, $\Sinside$ in the SCC and $\Soutside$ outside the SCC, as per the definition given in Section~\ref{sec:classification}. 
We append a degenerated leaf SCC as follows:

\begin{definition}[Appending a degenerated leaf SCC] \label{def:appending-degenerated}
To append a degenerated leaf SCC, first pick any vertex in $\Sinside$, say $v_\text{inside}$. 
We then add an arc according to the following two possibilities:
\begin{itemize}
\item Case 1: If $\Soutside$ contains exactly one non-leaf vertex, add an arc from $v_\text{inside}$ to the non-leaf vertex.
\item Case 2: Otherwise (all vertices in $\Soutside$ are leaf vertices), add an arc from $v_\text{inside}$ to an arbitrarily chosen vertex in $\Soutside$.
\end{itemize}
\end{definition}

\begin{proposition} \label{prop:degenerated}
After appending a degenerated leaf SCC, we have
\ifx\doublecolumn\undefined
 \begin{numcases}{\hspace{6.9ex}\total(\mathcal{G}') =}
\total(\mathcal{G}), \text{ or} \label{eq:degenerated-case-2a} \\
\total(\mathcal{G}) - 1, \label{eq:degenerated-case-2}
\end{numcases}
\vspace{-3ex}
\begin{align}
\tilde{\ell}^*(\mathcal{G}',\mathcal{U}') &= \tilde{\ell}^*(\mathcal{G},\mathcal{U}), \label{eq:decrease-opt-length-2}\\
V_\text{out}(\mathcal{G}') &= V_\text{out}(\mathcal{G}). \label{eq:equal-v-out-2}
\end{align}
\end{proposition}
\else
 \begin{numcases}{\hspace{7.7ex}\total(\mathcal{G}') =}
\total(\mathcal{G}), \text{ or} \label{eq:degenerated-case-2a} \\
\total(\mathcal{G}) - 1, \label{eq:degenerated-case-2}
\end{numcases}
\vspace{-4ex}
\begin{align}
\tilde{\ell}^*(\mathcal{G}',\mathcal{U}') &= \tilde{\ell}^*(\mathcal{G},\mathcal{U}), \label{eq:decrease-opt-length-2}\\
V_\text{out}(\mathcal{G}') &= V_\text{out}(\mathcal{G}). \label{eq:equal-v-out-2}
\end{align}
\end{proposition}
\fi

\begin{IEEEproof}
See Appendix~\ref{appendex:prop-degenerated}.
\end{IEEEproof}

\subsubsection{Pruning a leaf SCC} \label{sec:prop3}

The pruning step in the multi-sender case is the same as the single-sender case. Since we assume that all the messages are binary, we can choose any vertex from which we prune its outgoing arcs. Formally, we have the following:

\begin{definition}[Pruning a leaf SCC] \label{def:pruning}
To prune a leaf SCC, with vertex set $\S$, we arbitrarily select a vertex $v \in \S$, and remove all outgoing arcs from $v$.
\end{definition}

\begin{proposition} \label{prop:pruning}
After pruning a leaf SCC, we have
\begin{align}
\total(\mathcal{G}') &= \total(\mathcal{G}) - 1, \label{eq:decrease-n-total-3}\\
\tilde{\ell}^*(\mathcal{G}',\mathcal{U}') &\leq \tilde{\ell}^*(\mathcal{G},\mathcal{U}), \label{eq:decrease-opt-length-3}\\
V_\text{out}(\mathcal{G}') &= V_\text{out}(\mathcal{G}) -1. \label{eq:equal-v-out-3}
\end{align}
\end{proposition}

\begin{IEEEproof}
The pruning step grounds all the vertices in the leaf SCC, and does not change the connectivity of other vertices outside the leaf SCC. This gives \eqref{eq:decrease-n-total-3}.

As removing arcs reduces decoding requirements, any index code for $(\mathcal{G},\mathcal{U})$ also satisfies the decoding requirement for $(\mathcal{G}',\mathcal{U}')$. So, we have \eqref{eq:decrease-opt-length-3}.

As we remove all outgoing arc from a non-leaf vertex, we have \eqref{eq:equal-v-out-3}.
\end{IEEEproof}

\ifx\doublecolumn\undefined
\begin{figure}[t]
\centering
\includegraphics[width=6cm]{mwrc-it-10}
\caption{An algorithm that recursively prune/append leaf SCCs to produce a grounded information-flow graph}
\label{fig:algo-loop}
\end{figure}
\else
\begin{figure}[t]
\centering
\includegraphics[width=6.5cm]{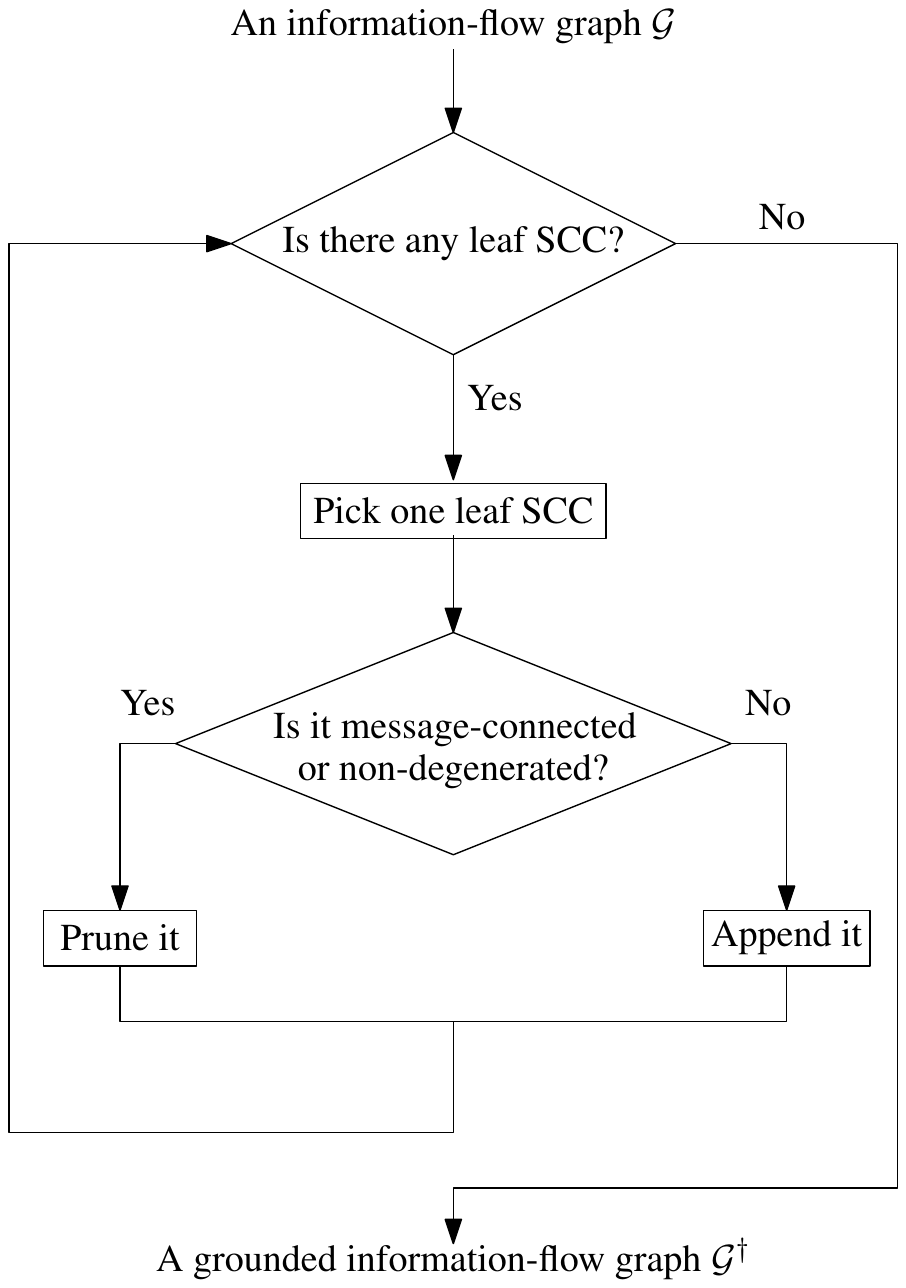}
\caption{An algorithm that recursively prune/append leaf SCCs to produce a grounded information-flow graph}
\label{fig:algo-loop}
\end{figure}
\fi

\subsection{Constructing an algorithm by pruning and appending}

We are now ready to assemble the pruning and appending steps to form an algorithm that returns a grounded information-flow graph $\mathcal{G}^\dagger$. 

From the previous section, we know that each step increases neither the optimal index-codelength nor the number of leaf SCCs. In fact, we have the following lemma:

\begin{lemma}
We can always get a grounded information-flow graph after a finite number of pruning and appending steps on all the leaf SCCs.
\end{lemma}
\begin{IEEEproof}
Consider an algorithm that recursively (i) appends a message-disconnected or a degenerated leaf SCC, and (ii) prunes a message-connected or a non-degenerated leaf SCC. We now show that the algorithm always terminates after a finite number of steps.

From Propositions~\ref{prop:disconnected}, \ref{prop:degenerated}, and \ref{prop:pruning}, the number of leaf SCCs always (i) decreases by one, or (ii) remains the same (In this case, the leaf SCC assimilates one or more vertices that do not belong to any leaf SCC. See Appendix~\ref{appendex:prop-degenerated} for proof.) after each pruning/appending step. The pruning/appending step that results in case (i) can be executed at most $n/2$ times as there are at most $n/2$ leaf SCCs in a graph. The pruning/appending step that results in case (ii) can be executed at most $n-2$ times, as this step requires a leaf SCC to start with (which must contain at least two vertices), and there are at most $n-2$ vertices that does not belong to any leaf SCC.

So, the algorithm will terminate after at most $3n/2 - 2$ appending/pruning steps. 
\end{IEEEproof}

Figure~\ref{fig:algo-loop} depicts our algorithm to find a lower bound on $\tilde{\ell}^*(\mathcal{G},\mathcal{U})$.

\subsection{Obtaining a lower bound}

The algorithm depicted in Figure~\ref{fig:algo-loop} returns possibly different lower bounds, $V_\text{out}(\mathcal{G}^\dagger)$, depending on which leaf SCC is selected in each iteration. We give an example to show this, after proving the following lemma:

\begin{lemma}
A non-degenerated leaf SCC may become degenerated after appending/pruning \textit{other} leaf SCCs.
\end{lemma}

\begin{IEEEproof}
Although no new leaf SCC is created after each appending/pruning step, more vertices may be grounded, and some non-degenerated leaf SCCs may become degenerated as the degeneration conditions are satisfied with more grounded vertices.  
\end{IEEEproof}

\begin{figure}[t]
\setlength{\baselineskip}{12pt}
\centering
\resizebox{25ex}{!}{%
\begin{tikzpicture}[node distance=8ex]
\tikzstyle{vertex}=[draw,circle]
\tikzstyle{connect}=[->,>=latex,bend right]
\tikzstyle{edge}=[color=red, ultra thick]
\node[vertex] (1) {1};
\node[vertex,below=of 1] (2) {2};
\node[vertex,right=of 1] (3) {3};
\node[vertex,right=of 2] (4) {4};
\node[vertex,right=of 3] (5) {5};
\node[vertex,right=of 4] (6) {6};
\path (1) edge [connect] (2)
(2) edge [connect] (1)
(3) edge [connect] (4)
(4) edge [connect] (3)
(5) edge [connect] (6)
(6) edge [connect] (5);
\path (1) edge [edge] (2)
(1) edge [edge] (3)
(3) edge [edge] (2)
(1) edge [edge] (4)
(2) edge [edge] (4);
\path (3) edge [edge] (5)
(3) edge [edge] (6)
(4) edge [edge] (5)
(4) edge [edge] (6);
\node[draw,dotted,fit=(1) (2),label={[align=left]below:message-\\connected}] (fit1) {};
\node[draw,dotted,fit=(3) (4),label={[align=left]below:non-\\degenerated}] (fit2) {};
\node[draw,dotted,fit=(5) (6),label={[align=left]below:degenerated}] (fit3) {};

\end{tikzpicture}
} %
\caption{An example showing the importance of the pruning/appending sequence. Arcs in $\mathcal{G}$ are marked with arrows, and edges in $\mathcal{U}$ with solid lines.}
\label{fig:order}
\end{figure}
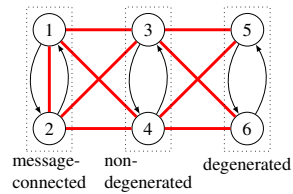

\begin{example}
Consider the index-coding instance depicted in Figure~\ref{fig:order}. The directed subgraph induced by vertices $\{1,2\}$ is a message-connected leaf SCC, that by $\{3,4\}$ non-degenerated, and that by $\{5,6\}$ degenerated. Suppose that we first prune the non-degenerated leaf SCC $\{3,4\}$, followed by pruning the message-connected $\{1,2\}$, and then appending the degenerated $\{5,6\}$. This pruning/appending sequence gives a resultant grounded graph with $V_\text{out}(\mathcal{G}^\dagger) = 4$. However, if we append $\{5,6\}$ and prune $\{1,2\}$ first, $\{3,4\}$ will be made degenerated and will be appended instead. This gives a different $\mathcal{G}^\dagger$ with a higher $V_\text{out}(\mathcal{G}^\dagger)=5$.
\end{example}

To obtain the tightest lower bound using the algorithm in Figure~\ref{fig:algo-loop}, we perform the following combinatorial optimization:
\begin{theorem}[Multi-sender: lower-bound] \label{theorem:combinatorial-lower-bound}
The optimal index-codelength for a multi-sender index-coding instance represented by $(\mathcal{G},\mathcal{U})$ is lower bounded as
\begin{equation} 
\tilde{\ell}^*(\mathcal{G},\mathcal{U}) \geq \max V_\text{out}(\mathcal{G}^\dagger), \label{eq:combinatorial-lower}
\end{equation}
where $\mathcal{G}^\dagger$ is the resultant graph after running the algorithm depicted in Figure~\ref{fig:algo-loop}, and the maximization is taken over all possible sequences of pruning/appending the leaf SCCs in the algorithm.
\end{theorem}

\ifx\doublecolumn\undefined
\begin{figure}[t]
\centering
\includegraphics[width=7cm]{mwrc-it-11}
\caption{An heuristic algorithm (which processes non-degenerated leaf SCCs last) that recursively prune/append leaf SCCs to produce a grounded information-flow graph}
\label{fig:algo-heuristic}
\end{figure}
\else
\begin{figure}[t]
\centering
\includegraphics[width=6cm]{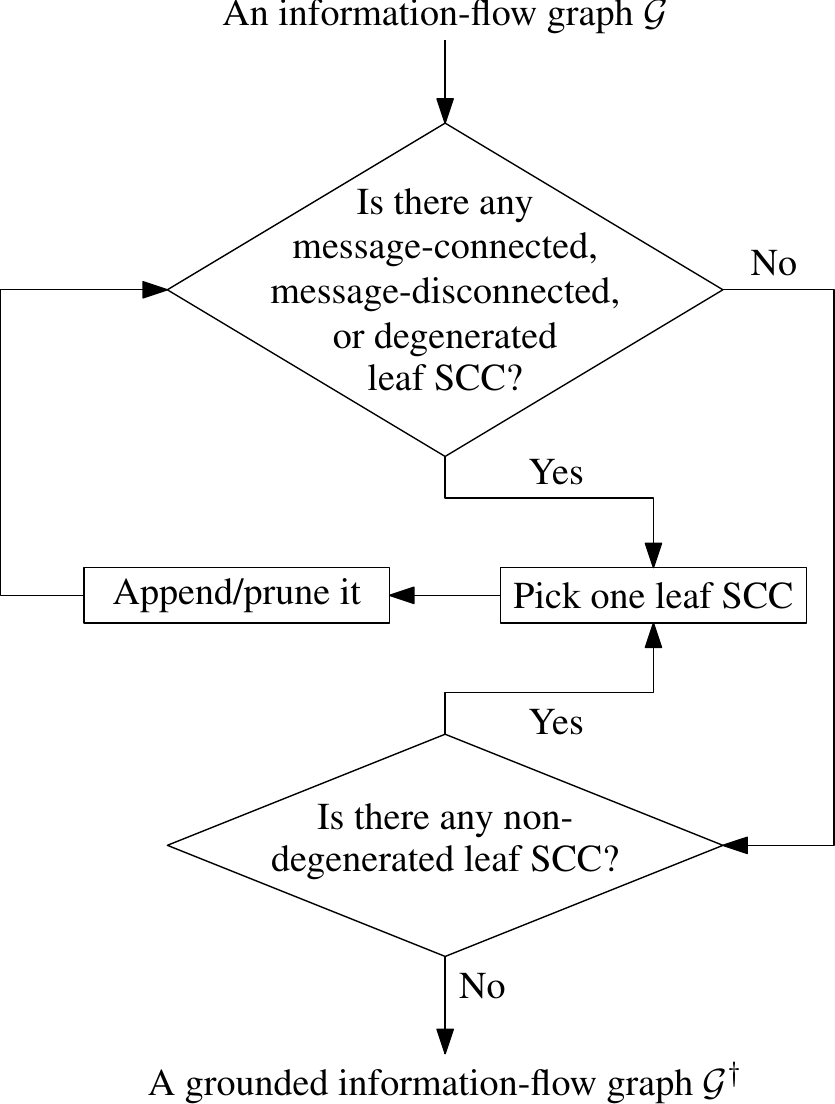}
\caption{An heuristic algorithm (which processes non-degenerated leaf SCCs last) that recursively prune/append leaf SCCs to produce a grounded information-flow graph}
\label{fig:algo-heuristic}
\end{figure}
\fi

\subsection{Another lower bound using a heuristic algorithm}

The lower bound obtained by the combinatorial optimization in Theorem~\ref{theorem:combinatorial-lower-bound} has some implementation issues:
\begin{itemize}
\item For a large number of leaf SCCs, the optimization runs in factorial time in the number of leaf SCCs.
\item It provides little insights to when the algorithm is optimal (i.e., giving a tight lower bound), and how good the algorithm is (whether we can bound the gap between the optimal index codelength and the lower bound).
\end{itemize}

In light of these issues, we design a heuristic algorithm. We first prove the following lemma:

\begin{lemma} \label{lemma:order-message}
The status of a leaf SCC being message-connected, message-disconnected, or degenerated is not affected by the appending/pruning of \textit{other} leaf SCCs.
\end{lemma}

\begin{IEEEproof}
See Appendix~\ref{appendix:order}.
\end{IEEEproof}

Lemma~\ref{lemma:order-message} suggests that it does not matter in what order we prune or append message-connected, message-disconnected, or degenerated leaf SCCs. At the end, all message-connected leaf SCCs will be pruned, and all message-disconnected and degenerated leaf SCCs appended.\footnote{Here, we do not optimize how to append a degenerated leaf SCC, when there are two or more ways to append it. So, we may get to a suboptimal solution.}

It matters, however, in which order we process \textit{non-degenerated} leaf SCCs. As pointed out in the previous section, having more grounded vertices increases the chances of a non-degenerated leaf SCC become degenerated for which we can append (instead of pruning it in the event that it stays non-degenerated). We will design a heuristic algorithm that leaves the \textit{pruning of the non-degenerated leaf SCCs to the last}---we ground the rest as far as possible.
This heuristic algorithm is shown in Figure~\ref{fig:algo-heuristic}.
Note that this heuristic algorithm is a one-pass algorithm which does not iterate over operating different sequences of leaf SCCs.

\begin{algorithm}[t]
\begin{small}
\SetKwInOut{Input}{input}
\SetKwInOut{Output}{output}

\Input{$(\mathcal{G},\mathcal{U})$}
\Output{$(\mathcal{G}^\dagger,\mathcal{U}^\dagger)$, where $\mathcal{G}^\dagger$ is grounded}
\BlankLine

\tcp{\scriptsize Initialization}
\ForEach{message-connected leaf SCC}{
Prune the message-connected leaf SCC\;
}
\While{there exists message-disconnected or degenerated leaf SCC}{
\ForEach{message-disconnected leaf SCC}{
Append the message-disconnected leaf SCC\;
}
\While{there exists degenerated leaf SCC}{
Append the degenerated leaf SCC\;
}
}
\tcp{\scriptsize Iteration}
\While{there exists leaf SCC}{
\tcp{\scriptsize Only message-connected or non-degenerated leaf SCCs left}
\If{there exists message-connected leaf SCC}{
Prune one message-connected leaf SCC\;
}
\Else(\tcp*[f]{\scriptsize Only non-degenerated leaf SCCs left}){
Prune one non-degenerated leaf SCC\;
}
\tcp{\scriptsize Remove all leaf SCCs that turn message-disconnected or degenerated}
\While{there exists message-disconnected or degenerated leaf SCC}{
\ForEach{message-disconnected leaf SCC}{
Append the message-disconnected leaf SCC\;
}
\While{there exists degenerated leaf SCC}{
Append the degenerated leaf SCC\;
}
}
}
\caption{Combined Appending-Pruning Algorithm}
\label{algorithm-multi-sender}
\end{small}
\end{algorithm}

While the heuristic algorithm in Figure~\ref{fig:algo-heuristic} produces a grounded $\mathcal{G}^\dagger$ that we require, we re-write the algorithm in a slightly different way, as Algorithm~\ref{algorithm-multi-sender}, which 
leads to a lower bound (see Theorem~\ref{theorem:multi-sender-lower-bound}) in a form similar to the upper bound to be derived in Theorem~\ref{theorem:multi-sender-achievable} later. Having similar forms, we will then identify classes of $(\mathcal{G},\mathcal{U})$ for which the bounds match.

We now describe the idea behind Algorithm~\ref{algorithm-multi-sender}.
Our lower bound, $V_\text{out}(\mathcal{G}^\dagger)$, relates to the number of non-leaf vertices in the original graph $\mathcal{G}$ via \eqref{eq:equal-v-out-1}, \eqref{eq:equal-v-out-2}, and \eqref{eq:equal-v-out-3}. We see that we only need to count the number of pruning steps. So, we design Algorithm~\ref{algorithm-multi-sender} such a way that we
\begin{itemize}
\item identify and count each pruning step,
\item separate the pruning of message-connected leaf SCCs in the original graph $\mathcal{G}$ out in the Initialization phase (so that we can express the lower bound as a function of $\mathcal{G}$),
\item process non-degenerated leaf SCCs last (after processing all other types of leaf SCCs).
\end{itemize}
The Initialization and Iteration phases are the same except that we only prune non-degenerated leaf SCCs in the latter, after pruning and counting all message-connected leaf SCCs in $\mathcal{G}$.

Note that any leaf SCC being operated is destroyed except for degenerated leaf SCCs (see Appendix~\ref{appendex:prop-degenerated}). In this case, the degenerated leaf SCC may assimilate other vertices to form a larger leaf SCC, which can be of any type, depending on how the new leaf SCC is connected in $\mathcal{U}$. The type of this newly-formed leaf SCC is to be identified during runtime. 

After running Algorithm~\ref{algorithm-multi-sender}, we obtain the following lower bound:
\begin{theorem}[Multi-sender: lower bound] \label{theorem:multi-sender-lower-bound}
The optimal index codelength for a multi-sender index-coding instance represented by $(\mathcal{G},\mathcal{U})$ is lower bounded as
\begin{subequations}
\begin{align}
\tilde{\ell}^*(\mathcal{G},\mathcal{U}) &\geq V_\text{out}(\mathcal{G}^\dagger) \label{eq:lower-bound-multi-sender}\\
 & = V_\text{out}(\mathcal{G}) - \#(\text{pruning steps in Initialization})\nonumber\\
&\quad - \#(\text{pruning steps in all Iterations}) \label{eq:count-connected}\\
& = V_\text{out}(\mathcal{G}) - (\connected(\mathcal{G},\mathcal{U}) + \iteration), \label{eq:count-iteration}
\end{align}
\end{subequations}
where $\connected(\mathcal{G},\mathcal{U})$ is the number of message-connected leaf SCCs in $\mathcal{G}$, and $\iteration$ is the number of times Iteration repeated in Algorithm~\ref{algorithm-multi-sender}.
\end{theorem}

\begin{IEEEproof}
The lower bound is obtained by running Algorithm~\ref{algorithm-multi-sender}.
Equation~\eqref{eq:lower-bound-multi-sender} follows from \eqref{eq:proposed-new-lower-bound}; \eqref{eq:count-connected} follows from \eqref{eq:equal-v-out-1}, \eqref{eq:equal-v-out-2}, and \eqref{eq:equal-v-out-3}; \eqref{eq:count-iteration} follows as one and only one leaf SCC is pruned during each Iteration.
\end{IEEEproof}


\begin{remark}
From \eqref{eq:count-iteration}, we note that the lower bound depends on the number of times Iteration is executed. 
In each iteration, the choice of which non-degenerated leaf SCC to prune will affect the lower bound.  As a rule of thumb, we prune a non-degenerated leaf SCC that will degenerate other non-degenerated leaf SCCs, so as to reduce the number of Iterations.
\end{remark}

\section{The Multi-Sender Case: An Upper Bound} \label{section:achievable}

We now derive an upper bound, via a coding scheme, by constructing special trees  in the message graph $\mathcal{U}$, referred to as \textit{connecting trees}. The vertex set of a connecting tree, denoted by $\mathcal{V}^\text{T}$, has the following properties:
\begin{enumerate}
\item Each vertex in $\mathcal{V}^\text{T}$ has one or more outgoing arcs in $\mathcal{G}$, and only to other vertices in $\mathcal{V}^\text{T}$. This means $\mathcal{V}^\text{T}$ has no outgoing arc to $\mathcal{V} \setminus \mathcal{V}^\text{T}$,


\item No vertex in $\mathcal{V}^\text{T}$ belongs to any message-connected leaf SCCs or another connecting tree.
\end{enumerate}
Let $\tree(\mathcal{G},\mathcal{U})$ denote the number of connecting trees that can be found for a given instance $(\mathcal{G},\mathcal{U})$. We will propose a coding scheme that achieves the following index codelength:

\begin{theorem}[Multi-sender: upper bound] \label{theorem:multi-sender-achievable}
The optimal index codelength for a multi-sender index-coding instance represented by $(\mathcal{G},\mathcal{U})$ is upper bounded as
\begin{equation}
\tilde{\ell}^*(\mathcal{G},\mathcal{U}) \leq V_\text{out}(\mathcal{G}) - ( \connected(\mathcal{G},\mathcal{U}) + \tree(\mathcal{G},\mathcal{U})).
\end{equation}
\end{theorem}

\ifx\doublecolumn\undefined
\begin{figure}[t]
\centering
\includegraphics[width=5.5cm]{mwrc-it-05}
\caption{An index-coding instance represented by $\mathcal{G}$ (with arrows) and $\mathcal{U}$ (with solid lines). There are two leaf SCCs in the graph: (i) message-connected, and (ii) degenerated. Here, we can form a connecting tree using the hollow vertices.}
\label{fig-2}
\end{figure}
\else
\begin{figure}[t]
\centering
\includegraphics[width=4.5cm]{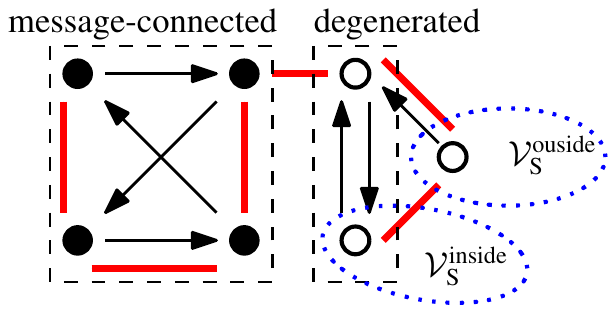}
\caption{An index-coding instance represented by $\mathcal{G}$ (with arcs) and $\mathcal{U}$ (with edges). There are two leaf SCCs in the graph: (i) message-connected, and (ii) degenerated. Here, we can form a connecting tree using the hollow vertices.}
\label{fig-2}
\end{figure}
\fi

The upper bound is optimized by finding the maximum number of connecting trees.

For example, consider the graphs in Figure~\ref{fig-2}. We can form one connecting tree using the hollow vertices.

\begin{IEEEproof}[Proof of Theorem~\ref{theorem:multi-sender-achievable}]
To simplify notation, we will show that there exists an index code of length $V_\text{out}- ( \connected + \tree)$, where we have dropped the arguments as they are clear from the context. 
Let a set of connecting trees be $\{\mathcal{T}_t = (\mathcal{V}^\text{T}_t, \mathcal{E}^\text{T}_t): t \in \{1,2,\dotsc, \tree\} \}$, and all the message-connected leaf SCCs in $\mathcal{G}$ be $\{\mathcal{C}_c = (\mathcal{V}^\text{C}_c, \mathcal{A}^\text{C}_c): c \in \{1,2,\dotsc, \connected\}\}$. Furthermore, let the remaining vertices in $\mathcal{G}$ be $\mathcal{V}' = \mathcal{V} \setminus \{ \bigcup_{t=1}^\tree \mathcal{V}^\text{T}_t \cup \bigcup_{c=1}^\connected \mathcal{V}^\text{C}_c \}$. Denote by $\mathcal{V}'_\text{out}$ the set of all non-leaf vertices in $\mathcal{V}'$. By definition, all $\mathcal{V}^\text{T}_t,\mathcal{V}^\text{C}_c,$ and $\mathcal{V}'$ are disjoint.


Our coding scheme is as follows:
\begin{enumerate}
\item For each connecting tree $(\mathcal{V}^\text{T}_t, \mathcal{E}^\text{T}_t)$, we transmit all $\{x_i \oplus x_j: (i,j) \in \mathcal{E}^\text{T}_t\}$, i.e., we transmit the XOR of the associated message pair for each edge. Note that we transmit $|\mathcal{V}^\text{T}_t|-1$ bits.
\item For each message-connected leaf SCC $(\mathcal{V}^\text{C}_c, \mathcal{A}^\text{C}_c)$ (which is edge-connected by definition), we first obtain a spanning tree in $\mathcal{U}$, denoted by $\mathcal{T}^\text{ST}_c = ( \mathcal{V}^\text{C}_c, \mathcal{E}^\text{ST}_c)$, where $\mathcal{E}^\text{ST}_c \subseteq \mathcal{E}$. We then transmit all $\{x_i \oplus x_j: (i,j) \in \mathcal{E}^\text{ST}_c\}$. Note that we transmit $|\mathcal{V}^\text{C}_c|-1$ bits.
\item For the rest of the non-leaf vertices, we transmit $\{ x_i: i \in \mathcal{V}'_\text{out}\}$, i.e., we transmit the message bits uncoded.
\end{enumerate}
Each vertex in the connecting trees and the message-connected SCCs has at least one outgoing arc. Hence, the coding scheme generates an index code of length $V_\text{out} - ( \connected + \tree)$.

We can easily verify that the index code can be transmitted, as each message pair to be XORed is associated with an edge, i.e., both the message bits belong to some sender.

Finally, we show that each receiver is able to obtain its requested messages. Recall that each receiver~$i$ needs to decode all messages in $\{ x_j: (j \rightarrow i) \in \mathcal{A}\}$. Now, each receiver~$i$ must belong to one---and only one---of the following groups:
\begin{enumerate}
\item (Connecting tree) $i \in \mathcal{V}^\text{T}_t$: Knowing $x_i$, receiver~$i$ can decode all $\{x_j: j \in \mathcal{V}^\text{T}_t\}$ from $\{x_j \oplus x_k: (j,k) \in \mathcal{E}^\text{T}_t\}$ by traversing the tree (which is connected by definition).
It can also decode the messages $\{x_k: k \in \mathcal{V}'_\text{out}\}$, sent uncoded.
Since all connecting trees and message-connected leaf SCCs have no outgoing arcs, each incoming arc to $i$ must be from either $\mathcal{V}^\text{T}_t \setminus \{i\}$ or $\mathcal{V}'_\text{out}$. So, receiver~$i$ is able to decode all its requested messages.
\item (Message-connected leaf SCC) $i \in \mathcal{V}^\text{C}_c$: Using the same argument as that for the connecting trees, we can show that receiver~$i$ can decode all its requested messages.
\item (The remaining vertices) $i \in \mathcal{V}'$: Using the argument in point 1, all incoming arcs to vertex $i$ must come from $\mathcal{V}'_\text{out} \setminus \{i\}$. Since we sent $\{x_j: j \in \mathcal{V}'_\text{out}\}$ uncoded, receiver~$i$ can decode all its requested messages. $\hfill\IEEEQEDhere$
\end{enumerate}
\end{IEEEproof}


\section{Special Cases Where The Bounds Are Tight and An Example}

Combining Theorems~\ref{theorem:multi-sender-lower-bound} and \ref{theorem:multi-sender-achievable}, we conclude $\iteration \geq \tree(\mathcal{G},\mathcal{U})$, and thus the optimal index codelength is found within $\iteration - \tree(\mathcal{G},\mathcal{U})$ bits. Recall that $\iteration$ is the number of times Iteration repeated in Algorithm~\ref{algorithm-multi-sender}, and it is upper bounded by the number of leaf SCCs in $\mathcal{G}$ that are not message-connected.
In the following special cases, we have $\iteration = \tree(\mathcal{G},\mathcal{U})$, and the lower bound is tight.




\begin{corollary} \label{corollary:no-semi}
If no leaf SCC remains after running the initialization of Algorithm~\ref{algorithm-multi-sender}, then
\begin{equation}
\tilde{\ell}^*(\mathcal{G},\mathcal{U}) = V_\text{out}(\mathcal{G}) - \connected(\mathcal{G},\mathcal{U}). \label{eq:unique}
\end{equation}
\end{corollary}

\begin{IEEEproof}
If Algorithm~\ref{algorithm-multi-sender} terminates after the initialization step, we have $\iteration = 0$, which implies that $\tree(\mathcal{G},\mathcal{U}) = \iteration = 0$.
\end{IEEEproof}

\begin{corollary} \label{corollary:unique}
If each bit $x_i$ in the message set $\mathcal{M}$ is known to only one sender (i.e., the $n$ sender constraint sets $\mathcal{M}_s$ partition $\mathcal{M}$),
then the optimal index codelength is given by \eqref{eq:unique}.
\end{corollary}

\begin{IEEEproof}
If messages $x_i$ and $x_j$ belong to some sender $s$ (i.e. $x_i, x_j \in \mathcal{M}_s$), then there exists an edge $(i,j)$ in the message graph $\mathcal{U}$.
Otherwise, if the messages $x_i,x_j$ belong to different senders, it is impossible to have a \textit{path} between $i$ and $j$.
This means we have only message-connected or -disconnected leaf SCCs, i.e.,
there is no semi leaf SCC. Thus, $\iteration=0$.
\end{IEEEproof}

Corollary~\ref{corollary:unique} includes the result of the single-sender instance~\cite{ongho12} as a special case.

Note that the scenario in Corollary~\ref{corollary:unique}, where the senders' message sets do not overlap, cannot be trivially solved by splitting the multi-sender instance into multiple single-sender instances, where each instance consists of the messages known to one sender. We illustrate this with an example in Appendix~\ref{appendix:multi-to-single}.

\ifx\doublecolumn\undefined
\begin{figure}[t]
\centering
\includegraphics[width=9.5cm]{mwrc-it-06a}
\caption{We run Algorithm~\ref{algorithm-multi-sender} on the graph in Figure~\ref{fig-2}. Subfigures~(a) and (b) show the initialization step where the message-connected leaf SCC is pruned (by removing the dashed arc in (a)) and the degenerated leaf SCC is appended (by adding the dashed arc in (b)), respectively. Subfigure~(c) shows the final graphs after the algorithm terminates.}
\label{fig-3}
\end{figure}
\else
\begin{figure}[t]
\centering
\includegraphics[width=8cm]{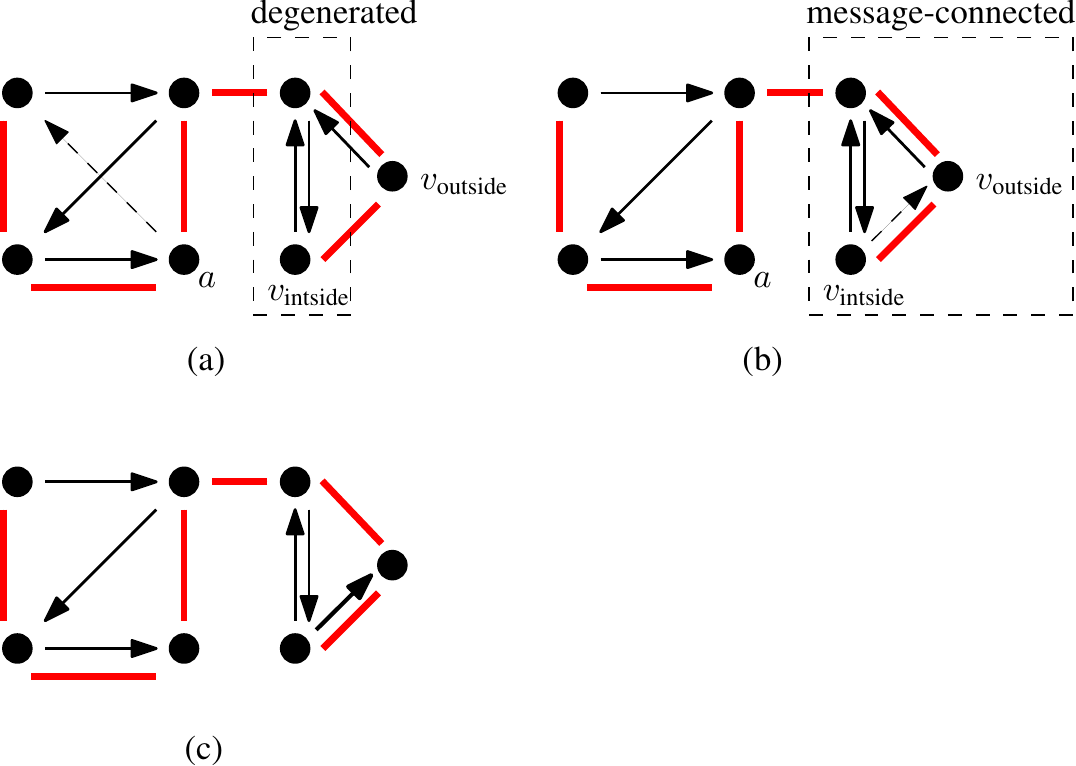}
\caption{We run Algorithm~\ref{algorithm-multi-sender} on the graph in Figure~\ref{fig-2}. Subfigures (a) and (b) show the initialization step where the message-connected leaf SCC is pruned (by removing the dashed arc in (a)) and the degenerated leaf SCC is appended (by adding the dashed arc in (b)), respectively. Subfigure (c) shows the final graphs after the algorithm terminates.}
\label{fig-3}
\end{figure}
\fi

\subsection{An example}

We now illustrate the lower bound and the achievability for the graph in Figure~\ref{fig-2}. For lower bound, we run Algorithm~\ref{algorithm-multi-sender} to obtain the resultant graph shown in Figure~\ref{fig-3}(c). Specifically, starting from Figure~\ref{fig-2}, note that the furthest-left four vertices form a message-connected leaf SCC. We prune this leaf SCC by removing all outdoing arcs from an arbitrarily selected vertex. This results in Figure~\ref{fig-3}(a), where we remove the outgoing arc (dashed arc) from $a$. Now, the two middle vertices forms a degenerated leaf SCC. We append this leaf SCC by adding an arc (indicated by the dashed arc in Figure~\ref{fig-3}(b)) from $v_\text{inside}$ to  $v_\text{outside}$. This completes the initialization step.

 Entering the iteration stage, we note that the right three vertices in Figure~\ref{fig-3}(b) now form a message-connected leaf SCC. We prune this leaf SCC by removing all outgoing arc(s) from the arbitrarily selected vertex $v_\text{outside}$. The algorithm terminates here, after destroying all leaf SCCs. From Theorem~\ref{theorem:multi-sender-lower-bound}, we have the lower bound $\tilde{\ell}^* \geq 7 - (1 + 1) = 5$.

For achievability, recall that we can form a connecting tree using the three hollow vertices in Figure~\ref{fig-2}. So, Theorem~\ref{theorem:multi-sender-achievable} gives the upper bound $\tilde{\ell}^* \leq 7 - (1 + 1) = 5$.

In this example, we have a scenario where $\iteration = \tree(\mathcal{G},\mathcal{U}) = 1$, and hence both the lower and upper bounds are tight, giving the optimal index codelength of five bits. This example illustrates that the upper and lower bounds can coincide even if $\iteration > 0$ (cf.\ Corollary~\ref{corollary:no-semi}).

\section{Discussions and Future Work}



We now show that the pairwise linear coding proposed in Section~\ref{section:achievable} can be suboptimal. Consider an index-coding instance with six messages and six receivers, with $\mathcal{A} = \{ (1 \leftrightarrow 2), (3 \leftrightarrow 4), (5 \leftrightarrow 6) \}$ where $(i \leftrightarrow j) \triangleq \{ (i \rightarrow j), (j \rightarrow i)\}$,
 and four senders having the following messages $(x_1,x_3,x_5)$, $(x_3,x_5,x_2)$, $(x_5,x_2,x_4)$, and $(x_2,x_4,x_6)$ respectively.

\ifx\doublecolumn\undefined
\begin{figure}[t]
\centering
\includegraphics[width=4cm]{mwrc-it-08}
\caption{An example where the optimal index codelength $\ell^*=4$ is achievable by XOR of three bits}
\label{fig:pairwise-suboptimal}
\end{figure}
\else
\begin{figure}[t]
\centering
\includegraphics[width=3cm]{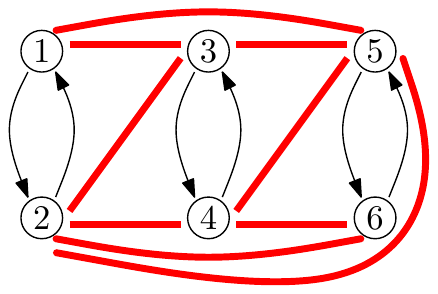}
\caption{An example where the optimal index codelength $\ell^*=4$ is achievable by XOR of three bits}
\label{fig:pairwise-suboptimal}
\end{figure}
\fi

The graphical representation is depicted in Figure~\ref{fig:pairwise-suboptimal}.
In this example, there is no message-connected, message-disconnected, or degenerated leaf SCC in $\mathcal{G}$.
Running Algorithm~\ref{algorithm-multi-sender},  
we get the lower bound $\tilde{\ell}^* \geq 4$.


We can also show that the largest $\tree = 1$, i.e., the pairwise linear coding (Theorem~\ref{theorem:multi-sender-achievable}) can only achieve 5 bits. However, if each sender sends the XOR of its three message bits, the 4-bit lower bound is achievable.

This example illustrates that---in stark contrast to the single-sender case---the pairwise coding scheme described in Section~\ref{section:achievable} is not always optimal. It also shows a disadvantage of using $\mathcal{U}$: it cannot differentiate a sender having $(x_1,x_2,x_3)$ from three senders having $(x_1,x_2), (x_2,x_3),$ and $(x_1,x_3)$ respectively. For future work, we will investigate a more general coding scheme and a more informative graphical representation.

\appendices

\section{A Graph Contains No Leaf SCC If and Only If It Is Grounded} \label{appendix:grounded}

\textit{Every directed graph with no leaf SCC is grounded:}
Given any graph $\mathcal{G}$, we form a \textit{supergraph} $\mathcal{G}^\text{s}$ by replacing each (leaf or non-leaf) SCC with at least two vertices  by a special vertex, referred to as a \textit{supernode}. First, $\mathcal{G}^\text{s}$ cannot contain any directed cycle.
Otherwise, all supernodes and vertices in the cycle form an SCC, and it would have been collapsed into a supernode.
Since $\mathcal{G}^\text{s}$ is acyclic, a path from any node/supernode must terminate somewhere.
Now, if the $\mathcal{G}$ has no leaf SCC, then $\mathcal{G}^\text{s}$ has no leaf supernode, and any path must terminate at a regular (i.e., non-supernode) leaf vertex.  $\hfill\blacksquare$ 

\textit{Every grounded directed graph has no leaf SCC:} This is easily seen by the fact that all vertices in the leaf SCC cannot be grounded, because these vertices can only reach vertices within the same leaf SCC, all of which are non-leaf. $\hfill\blacksquare$

\section{Proof of Proposition~\ref{prop:disconnected}} \label{appendex:proof-prop-disconnected}
The proof of Proposition~\ref{prop:disconnected} requires the following lemma. 
We defer the proof of this lemma to after that of the proposition.
\begin{lemma} \label{lemma:disconnected}
From any index code, any receiver is able to decode the messages of any message-disconnected leaf SCC.
\end{lemma}

\begin{IEEEproof}[Proof of Proposition~\ref{prop:disconnected}]
We first show that \eqref{eq:decrease-n-total-1} holds. 
The appending step adds a dummy leaf vertex and an arc from the message-connected leaf SCC to the dummy vertex. This grounds all vertices in the leaf SCC, and the leaf SCC is no longer a leaf SCC. The connectivity of all other vertices outside the leaf SCC is unchanged. So, this appending step neither creates any new leaf SCC nor destroys any other existing leaf SCC. 
Hence the number of leaf SCCs decreases by one after this step. We have \eqref{eq:decrease-n-total-1}.


Next, we show that \eqref{eq:decrease-opt-length-1} holds.
Adding the arc $(v \rightarrow n+1)$ now additionally requires the dummy receiver to decode messages in the appended leaf SCC, so $\tilde{\ell}^* ( \mathcal{G}', \mathcal{U}') \geq \tilde{\ell}^*(\mathcal{G},\mathcal{U})$.
But as the appended leaf SCC is message-disconnected, Lemma~\ref{lemma:disconnected} above says that its messages can be decoded without using any side-information. 
Hence, the dummy receiver can decode its required message defined in $\mathcal{G}'$ even using the index code for the original graph $(\mathcal{G},\mathcal{U})$.
We have \eqref{eq:decrease-opt-length-1}.

Finally we show \eqref{eq:equal-v-out-1}. 
A new arc is added from $v \in \S$ to  a newly added dummy vertex $n+1$. Vertex $v$ already had at least one outgoing arc before the appending step, and $n+1$ is a leaf vertex. So,
we have \eqref{eq:equal-v-out-1}.
\end{IEEEproof}

It still remains to prove Lemma~\ref{lemma:disconnected}. 

\begin{IEEEproof}[Proof of Lemma~\ref{lemma:disconnected}]
Let $\S$ be the vertex set of a message-disconnected leaf SCC.
By definition, there are two vertices $a,b \in \S$ that are not connected by any path in $\mathcal{U}$. Let $\mathcal{V}_a$ be a set of all vertices connected to $a$, and $\mathcal{V}_b \triangleq \mathcal{V} \setminus \mathcal{V}_a$. 
This means $a \in \mathcal{V}_a$, $b \in \mathcal{V}_b$, and there is no edge across $\mathcal{V}_a$ and $\mathcal{V}_b$.

The lack of edge-connectivity (in $\mathcal{U}$) between $\mathcal{V}_a$ and $\mathcal{V}_b$ implies that any index code can be partitioned into two parts (or subcodes), $\boldsymbol{c} = (\boldsymbol{c}_a, \boldsymbol{c}_b)$, such that every bit in $\boldsymbol{c}_a$  depends on only $\{x_i: i \in \mathcal{V}_a\}$ and not on $\{x_j: j \in \mathcal{V}_b\}$, and vice versa. This means the random variables $X_a$ and $(\boldsymbol{C}_b,X_b)$ are independent, even given $\boldsymbol{C}_a$.\footnote{Recall that we use upper-case letters to denote the corresponding random variables.} In other words,
\begin{equation}
X_a - \boldsymbol{C}_a - (\boldsymbol{C}_b,X_b) \label{eq:markov}
\end{equation}
forms a Markov chain.

Intuitively, the parts $\boldsymbol{c}_a$ (which contains $x_a$) and $\boldsymbol{c}_b$ do not have any message bit in common.
Then it must be true that receiver~$a$ decodes $x_b$ using only the part $\boldsymbol{c}_b$, i.e., without using its prior message~$x_a$, which appears only in the other separate part $\boldsymbol{c}_a$ (its prior messages $x_a$ can help receiver~$a$ to decode messages only in $\boldsymbol{c}_a$).
Hence, if receiver~$a$ can decode $x_b$, so can any receiver---even one without any prior message. 

We now prove this formally.
Since $a$ and $b$ belong to an SCC, receiver~$a$ must decode $x_b$ from the index code $\boldsymbol{c}$ and its prior information $x_a$ (see Lemma~\ref{lemma:predecessor}). This means
\begin{subequations}
\begin{align}
&H(\boldsymbol{C}_a, \boldsymbol{C}_b, X_a, X_b) = H(\boldsymbol{C}_a, \boldsymbol{C}_b, X_a)\\
\Rightarrow\quad & H(\boldsymbol{C}_a, \boldsymbol{C}_b, X_b) + H(X_a |\boldsymbol{C}_a, \boldsymbol{C}_b, X_b) \notag \\
&\quad  = H(\boldsymbol{C}_a, \boldsymbol{C}_b) + H(X_a | \boldsymbol{C}_a, \boldsymbol{C}_b)  \\
\Rightarrow\quad & H(\boldsymbol{C}_a, \boldsymbol{C}_b, X_b) + H(X_a |\boldsymbol{C}_a) \notag \\
&\quad = H(\boldsymbol{C}_a, \boldsymbol{C}_b) + H(X_a |\boldsymbol{C}_a) \label{eq:1-disconnected} \\
\Rightarrow\quad & H(\boldsymbol{C}_a, \boldsymbol{C}_b, X_b)= H(\boldsymbol{C}_a, \boldsymbol{C}_b) \label{eq:2-disconnected}
\end{align}
\end{subequations}
Here, \eqref{eq:1-disconnected} is due to the Markov chain~\eqref{eq:markov}. Equation~\eqref{eq:2-disconnected} means that any receiver can obtain $x_b$ from the index code alone.
Consequently, any receiver can also decode the messages of all predecessors of $b$ (using the  argument in the proof of Lemma~\ref{lemma:predecessor}), which include all messages in the message-disconnected leaf SCC. We have Lemma~\ref{lemma:disconnected}.
\end{IEEEproof}

\section{Proof of Proposition~\ref{prop:degenerated}} \label{appendex:prop-degenerated}

Let the appended arc be $(v_\text{inside} \rightarrow v_\text{outside})$. 
Unlike appending a message-disconnected leaf SCC, here, we change the connectivity of some vertex outside the leaf SCC, i.e., $v_\text{outside}$, by creating an incoming arc to it.

The following lemma will be used to prove Proposition~\ref{prop:degenerated}. 
\begin{lemma}\label{lemma:v-out-decode-v-inside}
Consider the appending step in Definition~\ref{def:appending-degenerated} for a degenerated leaf SCC with some chosen $\Sinside$ and $v_\text{outside}$. From any index code for $(\mathcal{G},\mathcal{U})$ (the problem instance \textit{before} the appending step), receiver~$v_\text{outside}$ must be able to decode all messages in $\Sinside$.
\end{lemma}
\begin{IEEEproof}
By Lemmas~\ref{lemma:predecessor} and \ref{lemma:predecessor-all}, receiver~$v_\text{outside}$ can decode the messages of all neighbors of $\Sinside$, denoted as $\mathcal{N}(\Sinside)$, as each vertex in $\mathcal{N}(\Sinside)$ is either grounded, a predecessor of $v_\text{outside}$, $v_\text{outside}$ itself, or a combination of these conditions.

Choose any $a \in \S \setminus \Sinside$ and $b \in \Sinside$. receiver~$a$ must decode $x_b$, because there is a path from $b$ to $a$ by the definition of SCC. We will show that if receiver~$a$ can decode $x_b$, so can receiver~$v_\text{outside}$. Consequently, receiver~$v_\text{outside}$ can decode all messages of $\Sinside$.

There is no edge across $\Sinside$ and $\S \setminus \Sinside$ by definition, meaning that any index codebit cannot be a function of the messages from both the sets. So, we can partition any index code for $(\mathcal{G},\mathcal{U})$ into $\boldsymbol{c} = (\boldsymbol{c}_1, \boldsymbol{c}_2)$, where
$\boldsymbol{c}_1$ does not contain  any message of $\Sinside$, and
$\boldsymbol{c}_2$ contains  only the messages of $\Sinside$ and $\mathcal{N}(\Sinside)$.

Intuitively, any advantage in decoding $x_b$ that receiver~$a$ has over receiver~$v_\text{outside}$ is due to knowing $x_a$, but $x_a$ can help receiver~$a$ only in decoding the messages in $\boldsymbol{c}_1$, which receiver~$a$ can then use to decode the messages in $\boldsymbol{c}_2$ (which contain $x_b$).
As receiver~$v_\text{outside}$ is able to decode the prior messages of $\mathcal{N}(\Sinside)$, which contains all the overlap of messages in $\boldsymbol{c}_1$ and $\boldsymbol{c}_2$, receiver~$v_\text{outside}$  is as capable as receiver~$a$ in decoding $x_b$.

We now prove this formally. Recall that $\boldsymbol{C}_2$ is a function of only messages $(\boldsymbol{X}_{\Sinside},\boldsymbol{X}_{\mathcal{N}(\Sinside)})$, where $\boldsymbol{X}_{\mathcal{S}} \triangleq \{X_i: i \in \mathcal{S}\}$ for some set $\mathcal{S}$; $\boldsymbol{C}_1$ can be written as a function of $(\boldsymbol{X}_{\mathcal{N}(\Sinside)}, \boldsymbol{X}_{\remain})$, where $\remain = \mathcal{V} \setminus (\Sinside \cup \mathcal{N}(\Sinside))$. Clearly, we have
\begin{subequations}
\begin{align}
& I(\boldsymbol{X}_{\remain};\boldsymbol{X}_{\Sinside}|\boldsymbol{X}_{\mathcal{N}(\Sinside)}) = 0 \label{eq:001}\\
\Rightarrow\quad & H(\boldsymbol{X}_{\remain}|\boldsymbol{X}_{\mathcal{N}(\Sinside)}) \notag \\
&\quad  - H(\boldsymbol{X}_{\remain}|\boldsymbol{X}_{\mathcal{N}(\Sinside)},\boldsymbol{X}_{\Sinside})= 0 \\
\Rightarrow\quad & H(\boldsymbol{X}_{\remain}|\boldsymbol{X}_{\mathcal{N}(\Sinside)},\boldsymbol{C}_2) \notag \\
&\quad  - H(\boldsymbol{X}_{\remain}|\boldsymbol{X}_{\mathcal{N}(\Sinside)},\boldsymbol{X}_{\Sinside}, \boldsymbol{C}_2) \leq 0 \label{eq:002}\\
\Rightarrow\quad & I(\boldsymbol{X}_{\remain};\boldsymbol{X}_{\Sinside}|\boldsymbol{X}_{\mathcal{N}(\Sinside)},\boldsymbol{C}_2) = 0 \label{eq:003} \\
\Rightarrow\quad & H(\boldsymbol{X}_{\Sinside}|\boldsymbol{X}_{\mathcal{N}(\Sinside)},\boldsymbol{C}_2) \notag \\
&\quad  - H(\boldsymbol{X}_{\Sinside}|\boldsymbol{X}_{\mathcal{N}(\Sinside)},\boldsymbol{X}_{\remain},\boldsymbol{C}_2) = 0 \\
\Rightarrow\quad & H(\boldsymbol{X}_{\Sinside}|\boldsymbol{X}_{\mathcal{N}(\Sinside)},\boldsymbol{C}_2,\boldsymbol{C}_1) \notag \\
&\quad  - H(\boldsymbol{X}_{\Sinside}|\boldsymbol{X}_{\mathcal{N}(\Sinside)},\boldsymbol{X}_{\remain},\boldsymbol{C}_2,\boldsymbol{C}_1) \leq 0 \label{eq:004} \\
\Rightarrow\quad & I(\boldsymbol{X}_{\remain};\boldsymbol{X}_{\Sinside}|\boldsymbol{X}_{\mathcal{N}(\Sinside)},\boldsymbol{C}_2,\boldsymbol{C}_1) = 0. \label{eq:005}
\end{align}
\end{subequations}
Here,\\
\eqref{eq:001} is due to the independence of the messages,\\
\eqref{eq:002} is derived because conditioning cannot increase entropy, and $\boldsymbol{C}_2$ is a function of only messages $(\boldsymbol{X}_{\Sinside},\boldsymbol{X}_{\mathcal{N}(\Sinside)})$,\\
\eqref{eq:003} and \eqref{eq:005} follow from the non-negativity of conditional mutual information,\\
\eqref{eq:004} is derived because conditioning cannot increase entropy, and $\boldsymbol{C}_1$ is a function of $(\boldsymbol{X}_{\mathcal{N}(\Sinside)},\boldsymbol{X}_{\remain})$.

Equation~\eqref{eq:005} implies that 
\begin{equation}
\boldsymbol{X}_{\remain} - (\boldsymbol{X}_{\mathcal{N}(\Sinside)},\boldsymbol{C}_2,\boldsymbol{C}_1) - \boldsymbol{X}_{\Sinside}
\end{equation}
forms a Markov chain. Since $a \in \remain$ and $b \in \Sinside$, 
\begin{equation}
X_a - (\boldsymbol{X}_{\mathcal{N}(\Sinside)},\boldsymbol{C}_2,\boldsymbol{C}_1) - X_b \label{eq:degen-markov}
\end{equation}
also forms a Markov chain.

Since $a$ and $b$ are both in the leaf SCC, receiver~$a$ must be able to decode $x_b$ from the index code and its prior information $x_a$, i.e., 
\begin{subequations}
\begin{align}
&& H(X_b | \boldsymbol{C}_1,\boldsymbol{C}_2,X_a) &= 0 \\
\Rightarrow && H(X_b |\boldsymbol{C}_1,\boldsymbol{C}_2,  X_a, X_{v_\text{outside}}, \boldsymbol{X}_{\mathcal{N}(\Sinside)}) &= 0  \label{eq:1-degen}\\
\Rightarrow && H(X_b | \boldsymbol{C}_1,\boldsymbol{C}_2, X_{v_\text{outside}}, \boldsymbol{X}_{\mathcal{N}(\Sinside)}) &= 0 \label{eq:3-degen}\\
\Rightarrow && H(X_b | \boldsymbol{C}_1,\boldsymbol{C}_2, X_{v_\text{outside}}) &= 0. \label{eq:2-degen}
\end{align}
\end{subequations}
Here,\\
 \eqref{eq:1-degen} is derived because conditioning cannot increase entropy,\\
\eqref{eq:3-degen} is due to the Markov chain~\eqref{eq:degen-markov},\\
\eqref{eq:2-degen} is due to $\boldsymbol{X}_{\mathcal{N}(\Sinside)}$ being a deterministic function of $(\boldsymbol{C}_1,\boldsymbol{C}_2, X_{v_\text{outside}})$ because receiver~$v_\text{outside}$ can decode $\boldsymbol{X}_{\mathcal{N}(\Sinside)}$ (which are the messages of the predecessors of vertex~$v_\text{outside}$).

Equation~\eqref{eq:2-degen} implies that knowing $x_{v_\text{outside}}$, receiver~$v_{\text{outside}}$ can decode $x_b$. Using the argument in the proof of Lemma~\ref{lemma:predecessor}, receiver~$v_{\text{outside}}$ must be able to decode all messages in the leaf SCC.
\end{IEEEproof}

\begin{IEEEproof}[Proof of Proposition~\ref{prop:degenerated}]
First, we show that either \eqref{eq:degenerated-case-2a} or \eqref{eq:degenerated-case-2} must hold.

Consider Case~2 in Definition~\ref{def:appending-degenerated}.
Similar to the proof of Proposition~\ref{prop:disconnected}, adding the arc causes the original leaf SCC $\mathcal{V}_S$ to become non-leaf. Furthermore, all other leaf SCCs do not change their status, since $v_\text{outside}$ is a leaf vertex and cannot belong to any leaf SCC. 
So, we get \eqref{eq:degenerated-case-2}. 

For Case 1 in Definition~\ref{def:appending-degenerated}, there are three possibilities: $v_\text{outside}$ is
\begin{itemize}
    \item Grounded: This case is similar to Case 2 above. The above argument holds even if $v_\text{outside}$ is not a leaf vertex but is grounded. Thus \eqref{eq:degenerated-case-2} holds.

\item Not grounded and has no directed path to $v_\text{inside}$: 
First, we show there is neither any destruction of existing SCCs, nor creation of any new one.
This follows because no cycle is created by the addition of the arc $(v_\text{inside} \rightarrow v_\text{outside})$, since $v_\text{outside}$ has no directed path back to $v_\text{inside}$.
Next, we need to consider if the appended arc may change some SCCs from leaf to non-leaf or vice versa.
For the leaf SCC $\S$, the appended arc makes it become non-leaf, since this is an outgoing arc which does not form a cycle. 
For SCCs other than $\S$, the appended arc is an incoming arc, which will not change the SCCs' being leaf or non-leaf. 
So, we have \eqref{eq:degenerated-case-2} for this particular case.

\item Not grounded and has a directed path to $v_\text{inside}$: 
Here, the appended arc creates a cycle between $v_\text{inside}$ (in $\S$) and $v_\text{outside}$ (outside $\S$).
This causes the original leaf SCC $\S$ to become a larger SCC by assimilating some external nodes (that include $v_\text{outside}$). 
If we could show that none of the assimilated external nodes belongs to any leaf SCC (meaning that no other leaf SCC is affected by this appending step), then clearly either \eqref{eq:degenerated-case-2a} or \eqref{eq:degenerated-case-2} holds depending on whether the enlarged SCC have outgoing arcs or not.
Indeed none of the assimilated external nodes belong to any leaf SCC, because they each have a directed path to $v_\text{inside}$.
Thus, we conclude the argument for this case.
\end{itemize}


To show \eqref{eq:decrease-opt-length-2}, we follow the same line of argument as in the proof of Proposition~\ref{prop:disconnected}. 
Let $(\mathcal{G},\mathcal{U})$ and $(\mathcal{G}',\mathcal{U}')$ denote the graphs before and after the appending step, respectively.
First, adding an arc increases the decoding requirements and result in $\tilde{\ell}^*(\mathcal{G}',\mathcal{U}') \geq \tilde{\ell}^*(\mathcal{G},\mathcal{U})$.
On the other hand, Lemma~\ref{lemma:v-out-decode-v-inside} says that receiver~$v_\text{outside}$ can decode all messages of $\Sinside$ using any index code for $(\mathcal{G},\mathcal{U})$. 
This implies that any index code for $(\mathcal{G},\mathcal{U})$ is also an index code for $(\mathcal{G}',\mathcal{U}')$. 
So, $\tilde{\ell}^*(\mathcal{G},\mathcal{U})$ is achievable for the problem instance $(\mathcal{G}',\mathcal{U}')$. 
So, \eqref{eq:decrease-opt-length-2} must hold.

Finally, as $v_\text{inside}$ already has at least one outgoing arc before the appending step (since it belongs to an SCC), this step does not change the number of vertices with outgoing arcs. We have \eqref{eq:equal-v-out-2}.
\end{IEEEproof}

\section{Proof of Lemma~\ref{lemma:order-message}} \label{appendix:order}
From Sections~\ref{sec:prop1} to \ref{sec:prop3} and Appendices~\ref{appendex:proof-prop-disconnected} to \ref{appendex:prop-degenerated}, we see that, besides the leaf SCC being operated on, all other leaf SCCs remain as leaf SCCs. In addition, pruning and appending neither remove nor add any edge in $\mathcal{U}$, and thus all other message-connected and message-disconnected leaf SCCs remain as leaf SCCs of the same type. For a degenerated leaf SCC not being operated on, it remains degenerated because we have the following after the pruning/appending operation on another leaf SCC:
\begin{itemize}
\item There is no addition or removal of edges in $\mathcal{U}$, and so $\Sinside$ remains unchanged, and the leaf SCC will neither become message-connected nor message-disconnected.
\item The number of non-leaf vertices in $\Soutside$ can only decrease, as all leaf vertices remain leaf vertices, and some non-leaf vertices may become leaf vertices (we might add more leaf vertices into $\Soutside$, see the argument below); and
\item Each neighbor of $\Sinside$ can still be in $\Soutside$ or be a predecessor of some vertex in $\Soutside$. This is because appending other leaf SCCs will not change the neighbor set. If we prune some other leaf SCC by removing all outgoing arcs from a vertex, say $v$, that happens to be in the path from a neighbor of $\Sinside$ to $\Soutside$, we just add $v$ (which has been made a leaf vertex) to $\Soutside$.
\end{itemize}

\section{A Multi-Sender Instance Cannot be Trivially Split into Multiple Single-Sender Instances} \label{appendix:multi-to-single}

Consider the following information-flow graph:
\begin{center}
\resizebox{11ex}{!}{%
\begin{tikzpicture}
\graph {[grow right=2cm, branch down=2cm, nodes={draw,circle},edge={>=latex}]
1 -> 3,
2 <- 4,
1 ->[bend left] 2,
2 ->[bend left] 1,
3 ->[bend left] 4,
4 ->[bend left] 3,
};
\end{tikzpicture}
}
\end{center}

Suppose that there are two senders, one having $(x_1, x_2)$, and the other $(x_3, x_4)$. We have a message-disconnected leaf SCC, and the optimal index-codelength is $\tilde{\ell}^* = 4$.

Consider splitting this index-coding instance into two single-sender instances according to the senders' messages. The first instance consists of the first sender with messages $(x_1, x_2)$. Since there are only two messages, we cannot have vertices~3 and 4 in the information-flow graph, i.e., we have 
\begin{center}
\resizebox{4ex}{!}{%
\begin{tikzpicture}
\graph {[grow right=2cm, branch down=2cm, nodes={draw,circle},edge={>=latex}]
1,2,
1 ->[bend left] 2,
2 ->[bend left] 1,
};
\end{tikzpicture}
}
\end{center}
This first single-sender instance has an optimal index codelength of $\ell^*=1$, and an optimal code of $x_1 \oplus x_2$.

The second instance can be similarly defined: a sender having $(x_3,x_4)$ and the information-flow graph is as follows:
\begin{center}
\resizebox{4ex}{!}{%
\begin{tikzpicture}
\graph {[grow right=2cm, branch down=2cm, nodes={draw,circle},edge={>=latex}]
3,4,
3 ->[bend left] 4,
4 ->[bend left] 3,
};
\end{tikzpicture}
}
\end{center}
This second single-sender instance has an optimal index codelength of $\ell^*=1$, and an optimal code of $x_3 \oplus x_4$. Combining these two instances gives a total codelength of $\tilde{\ell}^*=2$ with a code $(x_1 \oplus x_2, x_3 \oplus x_4)$. However, this is not an index code for the original instance as receiver~2 cannot decode $x_4$. In conclusion, we cannot trivially split a multi-sender uniprior instance into separate single-sender uniprior instances. Having said that, one can do so for \textit{unicast} instances.



\end{document}